\documentclass[11pt]{article}
\usepackage{amssymb}
\usepackage{amsmath}
\usepackage{amstext}
\usepackage{graphicx,epsfig}
\usepackage{epsfig}
\usepackage{verbatim} 
\usepackage{fancybox}
\usepackage{color}
\usepackage{ulem}
\usepackage{enumitem}
\usepackage{subfigure}
\usepackage{bbm}
\usepackage{parskip}
\usepackage[numbers,sort&compress]{natbib}
\usepackage{ytableau}

\newcommand{\Comment}[1]{{}}
\definecolor{darkblue}{rgb}{0.15,0.35,0.55}
\definecolor{reddish}{rgb}{0.65, 0.2, 0.2}
\usepackage[linktocpage=true]{hyperref}
\hypersetup{
colorlinks=true,
citecolor=darkblue,
linkcolor=reddish,
urlcolor=darkblue,
pdfauthor={},
pdftitle={},
pdfsubject={}
}

\newcommand{\be}{\begin{equation}}
\newcommand{\ee}{\end{equation}}
\newcommand{\bea}{\begin{eqnarray}}
\newcommand{\eea}{\end{eqnarray}}
\newcommand{\beas}{\begin{eqnarray*}}
\newcommand{\eeas}{\end{eqnarray*}}
\newcommand{\nn}{\nonumber}

\def\({\left(}
\def\){\right)}

\newcommand{\rd}{{\rm d}}

\def\gsim{ \lower .75ex \hbox{$\sim$} \llap{\raise .27ex \hbox{$>$}} }
\def\lsim{ \lower .75ex \hbox{$\sim$} \llap{\raise .27ex \hbox{$<$}} }

\def\xyma{\xymatrix@M.7em}
\def\xymas{\xymatrix@M.1em}
\newcommand{\ba}{\begin{eqnarray}}
\newcommand{\ea}{\end{eqnarray}}

\title{}
\author{}

\numberwithin{equation}{section}

\begin{document}
%
\renewcommand{\thefootnote}{\fnsymbol{footnote}}
~
\vspace{1.75truecm}
\begin{center}
{\LARGE \bf{Inflation in Flatland}}
\end{center} 

\vspace{1truecm}
\thispagestyle{empty}
\centerline{{\Large Kurt Hinterbichler,${}^{\rm a,}$\footnote{\href{mailto:kurt.hinterbichler@case.edu}{\texttt{kurt.hinterbichler@case.edu}}} Austin Joyce,${}^{\rm b,c,}$\footnote{\href{mailto:ajoy@uchicago.edu}{\texttt{austin.joyce@columbia.edu}}} and Justin Khoury${}^{\rm d,}$\footnote{\href{mailto:jkhoury@sas.upenn.edu}{\texttt{jkhoury@sas.upenn.edu}}}}}
\vspace{.5cm}
 
\centerline{{\it ${}^{\rm a}$CERCA, Department of Physics,}}
 \centerline{{\it Case Western Reserve University, 10900 Euclid Ave, Cleveland, OH 44106}} 
 \vspace{.25cm}
 
 \centerline{{\it ${}^{\rm b}$Center for Theoretical Physics, Department of Physics,}}
 \centerline{{\it Columbia University, New York, NY 10027}} 
 \vspace{.25cm}

\centerline{\it ${}^{\rm c}$Enrico Fermi Institute and Kavli Institute for Cosmological Physics,}
\centerline{\it University of Chicago, Chicago, IL 60637}
 \vspace{.25cm}

\centerline{\it $^{\rm d}$Center for Particle Cosmology, Department of Physics and Astronomy,}
\centerline{\it University of Pennsylvania, Philadelphia, PA 19104}

 \vspace{.8cm}
\begin{abstract}
\noindent
We investigate the symmetry structure of inflation in 2+1 dimensions. In particular, we show that the asymptotic symmetries of three-dimensional de Sitter space are in one-to-one correspondence with cosmological adiabatic modes for the curvature perturbation.
In 2+1 dimensions, the asymptotic symmetry algebra is infinite-dimensional, given by two copies of the Virasoro algebra,
and can be traced to the conformal symmetries of the two-dimensional spatial slices of de Sitter.
We study the consequences of this infinite-dimensional symmetry for inflationary correlation functions, finding new soft theorems that hold only in 2+1 dimensions. 
Expanding the correlation functions as a power series in the soft momentum $q$, these relations constrain the traceless part of the tensorial coefficient at each order in $q$ in terms of a lower-point function. As a check, we verify that the ${\cal O}(q^2)$ identity is satisfied by inflationary correlation functions in the limit of small sound speed.
\end{abstract}

\newpage

\setcounter{tocdepth}{2}
\renewcommand*{\thefootnote}{\arabic{footnote}}
\setcounter{footnote}{0}

\section{Introduction}
As we learn from A Square~\cite{abbott1884flatland}, things are often simpler in lower-dimensional settings. For instance, gravity in 2+1 dimensions is far simpler than gravity in our world, primarily because it has no local degrees of freedom~\cite{Deser:1983tn}. However it is not completely trivial because there can still be global and boundary degrees of freedom, leading to nontrivial solutions, {\it e.g.}, the BTZ black hole~\cite{Banados:1992wn,Banados:1992gq}. This makes 2+1 dimensions a nice testing ground, in which we have gained a variety of insights into the nature of gravitational physics (for a review, see~\cite{Carlip:1995zj}).

We are therefore motivated to investigate the (2+1)-dimensional version of cosmological inflation. Just as with pure gravity, the dynamics of inflation is much simpler in three dimensions.\footnote{Some work has been done on lower-dimensional inflation in the past:~\cite{Martinec:2014uva,Moore:2014sia} considered the dynamics of eternal inflation and the properties of perturbations in $(1+1)$-dimensional inflation. The question of starting inflation and some aspects of perturbations in $2+1$ dimensions were explored in~\cite{Banks:1984np,Samiullah:1991qv}. In~\cite{Kaplan:2014dia}, the authors considered aspects of the effective theory of holographic renormalization of 2-dimensional field theories, which is related to three-dimensional inflation by analytic continuation.} Because there are no gravitational waves, the only propagating degrees of freedom are the inflaton, plus any spectator fields that may be present.

It has recently become clear that viewing inflation as a process of spontaneous symmetry breaking is a powerful approach. This is exemplified by the effective field theory of inflation~\cite{Creminelli:2006xe,Cheung:2007st}, in which inflation is seen as a spontaneous breaking of time diffeomorphism invariance. 
It is also useful to think of gauge-fixed inflation as a process of global symmetry breaking, where the curvature perturbation, $\zeta$, is the corresponding Nambu--Goldstone mode~\cite{Creminelli:2012ed,Hinterbichler:2012nm,Assassi:2012zq}. The nonlinearly realized symmetries of inflation then lead to Maldacena's consistency relation~\cite{Maldacena:2002vr,Creminelli:2004yq} as a Ward identity~\cite{Assassi:2012zq,Hinterbichler:2013dpa,Kehagias:2012pd,Goldberger:2013rsa}. Including tensor perturbations, one finds an infinite set of soft theorems as Ward identities of an infinite number of non-linearly realized global symmetries~\cite{Hinterbichler:2013dpa}. These identities are the cosmological analogue of Adler's zero for soft pions~\cite{Adler:1964um}. The consistency relations have been generalized and extended in many directions: for example, they have been derived from CFT arguments~\cite{Schalm:2012pi,Mata:2012bx,McFadden:2014nta}, from the underlying diffeomorphism invariance~\cite{Berezhiani:2013ewa,Pimentel:2013gza,Collins:2014fwa,Armendariz-Picon:2014xda,Berezhiani:2014kga,Berezhiani:2014tda,Ferreira:2016hee}, extended to multiple soft momenta~\cite{Joyce:2014aqa,Mirbabayi:2014zpa}, and generalized to large scale structure~\cite{Kehagias:2013yd,Peloso:2013zw,Creminelli:2013mca,Creminelli:2013poa,Creminelli:2013nua,Horn:2014rta,Horn:2015dra,Hui:2016ffo}.

At their core, the single field soft theorems follow from the fact that $\zeta$-gauge does not fully fix the diffeomorphism symmetry of Einstein gravity. 
Rather, there remain residual large gauge transformations -- corresponding to conformal transformations of the spatial ${\mathbb R}^3$ slices -- under which $\zeta$ shifts nonlinearly. 
In $(2+1)$-dimensional inflation, the relevant group of symmetries is the conformal group of ${\mathbb R}^2$. 
This group is infinite-dimensional -- it is the Virasoro group familiar from two dimensional conformal field theory -- and correspondingly there is an infinite number of symmetries of $\zeta$.\footnote{In fact, in higher dimensions there is already an infinite number of residual large gauge transformation symmetries once one includes tensor modes and allows them to transform as well~\cite{Hinterbichler:2012nm,Hinterbichler:2013dpa,Bordin:2016ruc}. The difference in this case is that there is an infinite number of {\it scalar} symmetries.} 
One of our goals is to elucidate the nature of these symmetries and investigate how they act on fields during inflation.

Parallel to the investigation of soft limits in cosmology, there has been much recent interest in connecting soft theorems of quantum field theory on flat space to asymptotic symmetries ({\it e.g.},~\cite{Strominger:2013lka,Cachazo:2014fwa,Larkoski:2014bxa,Avery:2015rga}).
For example, Weinberg's soft graviton theorem~\cite{Weinberg:1965nx} can be thought of as a consequence of the Bondi--van der Burg--Metzner--Sachs (usually called BMS)~\cite{Bondi:1962px,Sachs:1962wk} invariance of gravitational scattering~\cite{Strominger:2013jfa,He:2014laa}. It seems natural that these two approaches should be closely connected. This connection has been made in one direction by~\cite{Mirbabayi:2016xvc}, who showed that the flat space identities for a U$(1)$ gauge theory can be derived from a generalization of Weinberg's construction of adiabatic modes in cosmology~\cite{Weinberg:2003sw}, and~\cite{Kehagias:2016zry}, who considered the cosmological analogue of the BMS group. Here we wish to further explore this connection in cosmology. We show that adiabatic modes are in one-to-one correspondence with the asymptotic symmetries of dS$_3$, so that either can be derived from the other. Our hope is that this will give some insight into extending this connection to cosmology in higher dimensions \cite{Ferreira:2016hee}.

Three dimensional de Sitter space has an infinite-dimensional asymptotic symmetry group~\cite{Brown:1986nw,Strominger:1997eq}. This can be thought of as a manifestation of the putative dS/CFT correspondence~\cite{Strominger:1997eq}, where theories on dS$_3$ have dual descriptions in terms of a 2d CFT, with corresponding Virasoro symmetry. In Sec.~\ref{sec:asympsymm} we review the derivation of the asymptotic symmetry transformations of dS$_3$ and how they close to form two copies of the Virasoro algebra. We then derive how these symmetries act on the curvature perturbation, $\zeta$. As anticipated, on the future boundary asymptotic symmetry transformations act as conformal transformations of ${\mathbb R}^2$. In Sec.~\ref{sec:zetasymms} we derive the same transformations for $\zeta$ by applying Weinberg's adiabatic mode construction to find residual large gauge transformations which may be smoothly connected to a physical solution. We find that the adiabatic modes for $\zeta$ specialized to de Sitter space reproduce precisely the asymptotic symmetry transformations with Brown--Henneaux boundary conditions.

Recall that in 3+1 dimensions the non-linearly realized symmetries, the dilation and special conformal transformations on ${\mathbb R}^3$, induce respectively a constant and linear gradient profile for~$\zeta$. In 2+1 dimensions, the additional Virasoro symmetries can induce profiles to all orders in the coordinates. Specifically, at each order in $n$, the holomorphic and anti-holomorphic symmetry generators can be arranged into a symmetric, traceless $n$-index tensor, which acts on $\zeta$ as
\be
\delta_{{i_1\cdots i_n}}\zeta \,\propto\,  x^{(i_1}\cdots x^{i_n)_{\rm T}}+\ldots\,,
\label{eq:tracelessintro}
\ee
where $x^{(i_1}\cdots x^{i_n)_{\rm T}}$ denotes the fully symmetric and traceless combination. These transformations are reminiscent of the tensor symmetries in 3+1 dimensions~\cite{Hinterbichler:2013dpa}.

In Sec.~\ref{sec:Ward} we derive the Ward identities associated to these symmetries. In momentum space they relate the symmetric, traceless part of the correlation function at each order $q^n$ in the soft momentum to a symmetry transformation on a lower-point function without the soft mode. We then go on to check some of these identities involving the 3-point function in Sec.~\ref{sec:check}. Computing the squeezed limit of the 3-point function analytically in odd space-time dimensions is somewhat tricky, because the mode functions are Hankel functions of integer order. Instead we calculate the squeezed limit in arbitrary even dimensions and continue the expression to $D=3$.  This requires constructing the EFT of inflation in arbitrary dimension and computing its quadratic and cubic action. We comment on future directions in Sec.~\ref{sec:concl}. Some technical appendices collect useful results.

\vspace{-.4cm}
\paragraph{Conventions:} We use the mostly-plus metric signature. We use $D$ to represent the space-time dimension and $d$ to represent the spatial dimension, hence $D = d+1$. In many cases it will be useful to work in terms of the complex coordinates $z = x+iy$ and $\bar z = x-iy$. Derivatives with respect to these coordinates are denoted by $\partial$ and $\bar{\partial}$. See Appendix~\ref{app:complexcoords} for more details.

\section{Asymptotic symmetries of dS${}_3$}
\label{sec:asympsymm}
We begin by reviewing the structure of the asymptotic symmetry algebra in three-dimensional de Sitter space. Our approach follows that of~\cite{Strominger:2001pn}, which adapted the seminal analysis of Brown and Henneaux~\cite{Brown:1986nw} to the dS$_3$ case.

We work in the planar slicing of de Sitter, using complex coordinates for the spatial slices and conformal time. The line element takes the form
\be
\rd s^2 = \frac{1}{H^2\eta^2}\left(-\rd\eta^2+\rd z\rd \bar z\right),\ \ \ ~~~~~~~ \eta\in (-\infty,0)\,,
\label{eq:dslineelement}
\ee
where the future boundary is at $\eta=0$.  

We search for diffeomorphisms that preserve the $\eta\rightarrow 0$ asymptotic structure of the metric associated to~\eqref{eq:dslineelement}. In order to do this, we have to specify some fall-off conditions that define what it means to preserve the asymptotics. In general, different choices of fall-off conditions will lead to different asymptotic symmetry algebras, and the choice of fall-off conditions is part of the data specifying the theory.\footnote{Choosing fall-off conditions is somewhat of an art: the goal is to choose sufficiently restrictive conditions such that only a subset of diffeomorphisms are allowed, but grant enough freedom for there to be interesting structure to the surviving transformations.} The conditions we demand are the analytic continuation of the Brown--Henneaux conditions~\cite{Brown:1986nw}
\begin{align}
\nonumber
g_{\eta\eta} &= -\frac{1}{H^2\eta^2}+{\cal O}(1)\,;~~~~~~~~g_{zz} = g_{\bar z\bar z}= {\cal O}(1)\,;\\
g_{z\bar z} &= \frac{1}{2H^2\eta^2}+{\cal O}(1)\,;~~~~~~~\,~g_{\eta z} = g_{\eta \bar z}= {\cal O}(1)\,.
\label{dsasymp2}
\end{align}
These fall-off conditions can be motivated by demanding that the charges associated to the Brown--York stress tensor~\cite{Brown:1992br} be finite and non-zero~\cite{Spradlin:2001pw}.

A general diffeomorphism acts on the metric via the Lie derivative: $\delta g_{\mu\nu} = \pounds_\xi g_{\mu\nu} = \xi^\alpha\partial_\alpha g_{\mu\nu}+g_{\alpha\mu}\partial_\nu\xi^\alpha+g_{\alpha\nu}\partial_\mu\xi^\alpha$. The vector fields which preserve the asymptotic conditions~\eqref{dsasymp2} can be parameterized in terms of two independent functions, one holomorphic $U(z)$ and one anti-holomorphic~$\bar U(\bar z)$.  
The most general vector field which preserves the asymptotic structure of dS$_3$ can be written as a linear combination of
\bea
&&\xi = \frac{\eta}{2}\partial U(z)\partial_\eta+ U(z)\partial + \frac{\eta^2}{2}\partial^2U(z)\bar\partial\,; \nn\\
&&\bar \xi = \frac{\eta}{2}\bar\partial\bar U(\bar z)\partial_\eta+ \frac{\eta^2}{2}\bar\partial^2\bar U(\bar z)\partial+ \bar U(\bar z)\bar\partial \,.
\label{eq:asympKillinV}
\eea
We expand the functions
 \be U(z) = -\sum_n a_nz^{n+1},\ \ \  \bar U(\bar z) = -\sum_n \bar a_n\bar z^{n+1}\, ,
 \ee and consider each of the modes
so that the vectors which generate the asymptotic symmetries are given by $\xi = \sum_n a_n\ell_n,\ \bar\xi = \sum_n \bar a_n\bar \ell_n$, with
\bea
&&\ell_n = -\frac{(n+1)}{2}z^{n}\eta\partial_\eta-z^{n+1}\partial-\frac{\eta^2}{2}n(n+1)z^{n-1}\bar\partial\,;\nn\\
&&\bar \ell_n = -\frac{(n+1)}{2}\bar z^{n}\eta\partial_\eta-\frac{\eta^2}{2}n(n+1)\bar z^{n-1}\partial-\bar z^{n+1}\bar\partial\,.
\label{eq:Lnvectors}
\eea
We will restrict our attention to the symmetries with $n\geq -1$, as these are the transformations which induce nonsingular profiles for the boundary stress tensor.
The Lie bracket between two of these transformations is given by
\be
[ \ell_n, \ell_m] = (n-m)\ell_{n+m}\,;~~~~~~~~~~~ [ \bar \ell_n, \bar \ell_m] = (n-m)\bar \ell_{n+m}\,.
\label{eq:wittalg}
\ee
Thus we see that the algebra of asymptotically Killing vectors is two copies of the Witt algebra.  If these symmetries are realized canonically, with charges $L_n$ and $\bar L_n$ corresponding to the symmetries generated by $\ell_n$ and $\bar \ell_n$, the Poisson bracket between charges develops a central extension. The asymptotic symmetry algebra then becomes two copies of the Virasoro algebra~\cite{Brown:1986nw}
\begin{align}
\nonumber
[ L_n, L_m] &= (n-m)L_{n+m} +\frac{c}{12}n(n^2-1)\delta_{n+m,0} \,; \\
[ \bar{L}_n, \bar{L}_m] &= (n-m)\bar{L}_{n+m}  +\frac{\bar c}{12}n(n^2-1) \delta_{n+m,0}\,.
\end{align}

The central charges $c$ and $\bar c$ can be extracted from the transformation properties of the boundary stress tensor, following~\cite{Balasubramanian:1999re}. A diffeomorphism corresponding to the vector~\eqref{eq:asympKillinV} shifts the line element~\eqref{eq:dslineelement} as follows
\be
\rd s^2 = \frac{1}{H^2\eta^2}\left(-\rd\eta^2+\rd z\rd \bar z\right)+\frac{1}{2H^2}\partial^3U(z)\rd z^2+\frac{1}{2H^2}\bar\partial^3\bar U(\bar z)\rd\bar z^2\,.
\label{eq:asympmetric}
\ee
This corresponds to a shift of the Brown--York stress tensor of pure dS$_3$ by\footnote{Note that the boundary stress tensor for a deformation of dS$_3$ of the form $\rd s^2 = \frac{1}{H^2\eta^2}(-\rd\eta^2 +\rd z\rd\bar z+h_{ij}\rd z^i\rd z^j)$, with $h$ transverse and traceless, is given by the leading piece of $h_{ij}$ near the boundary $T_{ij} = -\frac{1}{4GH\eta^2}h_{ij}\big\rvert_{\eta\to0}$~\cite{deHaro:2000vlm,Balasubramanian:2001nb}.
}
\be
\delta_U T(z) = -\frac{1}{8 GH}\partial^3 U\,;~~~~~~~~~~~\delta_{\bar U} \bar T(\bar z) = -\frac{1}{8 GH}\bar\partial^3 \bar U\,.
\ee
Using the CFT transformation rule for the stress tensor~\cite{Ginsparg:1988ui}
\be
\delta_U T = U \partial T+2 T\partial U+\frac{c}{12 }\partial^3 U\,;~~~~~~~~~~~\delta_{\bar U} \bar T = \bar U \bar \partial \bar T+2 \bar T\bar\partial\bar U+\frac{\bar c}{12 }\bar\partial^3\bar U\,,
\label{eq:cftTtrans}
\ee
and the fact that the stress tensor for pure dS$_3$ vanishes, this implies that the central charges are given by~\cite{Park:1998qk,Balasubramanian:2001nb}\footnote{Actually, there is a factor of $i$ due to the modes of the stress tensor being $iL_n$ rather than $L_n$, so that the central charge is imaginary. See~\cite{Balasubramanian:2001nb}.}
\be
c=\bar c=-\frac{3}{2 GH}\,.
\ee
The central charge will not play a major role in what follows, as we will need only the classical action of the asymptotic symmetries.

Our primary interest lies in understanding how these asymptotic symmetries act on the inflationary curvature perturbation $\zeta$. This variable is defined by the gauge choice where 
the scalar fluctuations are set to zero, $\delta\phi = 0$, and the spatial line element takes the form\footnote{For now the presence of the lapse and shift variables in the ADM parameterization is not important, but will play an important role in Section~\ref{sec:zetasymms}.}
\be
h_{ij}\rd z^i\rd z^j = a^2(\eta)e^{2\zeta(\eta,z,\bar z)}\rd z\rd\bar z\,.
\label{zetagaugespatial}
\ee
Thus $\zeta$ may be thought of as the Nambu--Goldstone boson corresponding to the spontaneously broken dS isometries during inflation~\cite{Creminelli:2012ed,Hinterbichler:2012nm}.
In order to derive the transformation rule for $\zeta$, 
consider a generalization of the metric~\eqref{eq:dslineelement} which induces a perturbed boundary metric (this is essentially the $\zeta$-gauge metric)
\be
\rd s^2 = \frac{1}{H^2\eta^2}\left(-\rd\eta^2+e^{2\zeta(\eta,z,\bar z)}\rd z\rd\bar z\right)\,.
\label{eq:pertrubedmetric}
\ee
The metric~\eqref{eq:pertrubedmetric} is not of the form~\eqref{eq:asympmetric} but can be put in this form if $\zeta = \zeta(z)$ is a function of $z$ alone. This is accomplished by first rescaling the time coordinate as $\eta \mapsto e^\zeta \eta$,
followed by the coordinate shift
$\bar z \mapsto \bar z+\eta^2\partial\zeta$,
after which the line element is given by
\be
\rd s^2 = \frac{1}{H^2\eta^2}\left(-\rd\eta^2+\rd z\rd \bar z\right)+\frac{1}{2H^2}\left(2\partial^2\zeta-2(\partial\zeta)^2\right)\rd z^2\,.
\label{eq:zzetametric}
\ee
This metric is clearly of the form~\eqref{eq:asympmetric}, and in fact solves the Einstein's equations everywhere~\cite{Banados:1998gg}.\footnote{Although we are working in the context of Einstein gravity throughout most of our discussion, the kinematic considerations of this Section actually rely only upon the asymptotic structure of the metric. Even in non-Einsteinian gravity theories, so long as they are fundamentally diffeomorphism invariant and possess asymptotically (A)dS solutions, the asymptotic symmetries will map solutions to solutions and all of our arguments should go through. Subleading terms in $\eta$ of~\eqref{eq:asympmetric} will likely be theory-dependent, however.}
Furthermore, we can induce this metric via an asymptotic symmetry transformation by matching\footnote{Note that the induced combination $-\partial^2\zeta + (\partial\zeta)^2 = R_{zz}$ is the $zz$ component of the Ricci tensor of the 2d surface at the future boundary.}
\be
2\partial^2\zeta - 2(\partial\zeta)^2 = \partial^3 U(z)\,.
\label{eq:inducedzeta}
\ee
If we then consider doing another asymptotic symmetry transformation on~\eqref{eq:zzetametric}, 
in order to reproduce the transformation rule~\eqref{eq:cftTtrans}, $\zeta$ must transform as
\be
\delta_U\zeta = \frac{1}{2}\partial U +U\partial\zeta\,.
\ee
As before, we can expand $U(z) = -\sum_n a_nz^{n+1}$, which leads to the transformation rule for $\zeta$~\eqref{eq:virzeta1}
\be
\delta_n\zeta = -\frac{1}{2}(n+1)z^n-z^{n+1}\partial \zeta\,,
\label{eq:virzeta1}
\ee
and similarly for the barred transformations coming from the $\bar U(\bar z)$ symmetries. 

Matching at linear order in $\zeta$, from~\eqref{eq:inducedzeta} we see that an asymptotic symmetry transformation $\ell_n$ induces a long-wavelength $\zeta$ profile of the form $\zeta\sim z^{n}$. This is the natural generalization of the conformal transformations in higher dimensions, which can induce a constant or linear profile for $\zeta$~\cite{Creminelli:2012ed,Hinterbichler:2012nm}, corresponding to our $n=0,1$ symmetries. Thus in 2+1 dimensions we can induce a profile to arbitrary order $n$.

This situation is somewhat analogous to spontaneous symmetry breaking. Transformation by an asymptotic symmetry maps the space-time to a different configuration, one which differs precisely by a soft $\zeta$ mode. Note that the asymptotic symmetries map solutions to the gravitational equations to other solutions with physical soft modes, which is precisely what the adiabatic mode construction achieves. Indeed, we will find that the constructions are actually equivalent.

It is worthwhile to make the connection between the symmetries in complex coordinates, $(z,\bar z)$ and the more familiar Cartesian coordinates $(x,y)$. The linear combinations $\delta_{-1}\pm\bar\delta_{-1}$ are related to translations of the Cartesian coordinates, while $\delta_0-\bar\delta_0$ corresponds to a 2d rotation. Furthermore, $\delta_0+\bar\delta_0$ generates a 2d dilation, while $\delta_{1}\pm\bar\delta_{1}$ generate 2d special conformal transformations. These last 3 symmetries are all realized nonlinearly on $\zeta$. All of these symmetries are familiar from the higher-dimensional case, where they act on $\zeta$ in a similar way. The first new symmetries are generated by $\delta_2,\bar\delta_2$; these can be combined in Cartesian coordinates into a rank-2 symmetric traceless tensor generator, which acts on $\zeta$ as 
\be
\delta_{{ij}}\zeta = x^{(i} x^{j)_{\rm T}} + \frac{2}{3}x^{(i} x^{j)_{\rm T}}x^k\partial_k\zeta- \frac{1}{3}\vec x^2x^{(i}\partial^{j)_{\rm T}}\zeta\,.
\label{eq:n2symmetry}
\ee
The transformation of $\zeta$ corresponds to a nonlinear shift by a traceless combination of the coordinates plus a piece linear in the field $\zeta$.  The nonlinear shift $\delta_{{ij}}\zeta\propto x^{(i} x^{j)_{\rm T}}$ is a spatial version of the enhanced symmetry of the special galileon \cite{Hinterbichler:2015pqa}, just as the nonlinear shift $\delta_{{i}}\zeta\propto x^{i}$ is a spatial version of the extended shift symmetry of the generic galileon \cite{Nicolis:2008in}.
The pattern continues for all the higher symmetries: at each order in $n$, the symmetry generators $\delta_n$ and $\bar{\delta}_n$ can be arranged into a symmetric, traceless $n$-index tensor, which acts on $\zeta$ as\footnote{The conformal Killing vectors which generate these symmetries are of the form
\be
\xi_i = \sum_{n=-1}^\infty \frac{1}{(n+2)!}M_{i j_1\cdots j_{n+1}}x^{j_1}\cdots x^{j_{n+1}},
\ee
where the coefficient tensors $M_{i j_1\cdots j_{n+1}}$ are symmetric and traceless in their last $n+1$ indices and satisfy the relation
\be
M_{(ij_1)j_2\cdots j_{n+1}} = \frac{1}{2}\delta_{ij_1}M^k_{kj_2\cdots j_{n+1}}.
\ee
}
\be
\delta_{{i_1\cdots i_n}}\zeta = x^{(i_1}\cdots x^{i_n)_{\rm T}}+\frac{2}{n+1} x^{(i_1}\cdots x^{i_n)_{\rm T}}x^k\partial_k\zeta-\frac{1}{n+1}\vec x^2 x^{(i_1}\cdots x^{i_{n-1}}\partial^{i_n)_{\rm T}}\zeta \,.
\label{eq:tracelesssymshift}
\ee
The form of these symmetries is quite similar to the infinite family of symmetries of $(3+1)$-dimensional inflation uncovered in~\cite{Hinterbichler:2013dpa}. Furthermore, the nonlinear shift $\delta_{{i_1\cdots i_n}}\zeta \,\propto\,  x^{(i_1}\cdots x^{i_n)_{\rm T}}$ is a spatial version of the extended shift symmetries of higher order galileons \cite{Hinterbichler:2014cwa,Griffin:2014bta}.

\section{Adiabatic modes and asymptotic symmetries}
\label{sec:zetasymms}
The approach we have taken in Sec.~\ref{sec:asympsymm} is somewhat unconventional from the point of view of cosmology.  In the cosmological context, to deduce the nonlinear transformation of $\zeta$ one typically looks for residual large gauge transformations which can be smoothly extended to physical solutions -- so-called adiabatic modes~\cite{Weinberg:2003sw}. In this Section we show that these two viewpoints are in fact equivalent. We will see explicitly that the asymptotic symmetry transformations are precisely the adiabatic modes of $\zeta$ in asymptotically de Sitter space.

Recall that the variable $\zeta$ is defined by the gauge choice  $\delta\phi = 0$ along with the spatial metric being of the form
\be
h_{ij} = e^{2\zeta}\left(e^\gamma\right)_{ij},
\ee
with $\gamma_{ij}$ transverse and traceless: $\partial_i\gamma^i_j = \gamma^i_i = 0$.\footnote{Note that there do not exist any transverse-traceless tensors which asymptote to zero at infinity in $(2+1)$-dimensions. However, we formally allow for the presence of graviton perturbations to account for the possibility that a profile is generated by the symmetries which does not fall off at infinity.}
We search for residual diffeomorphisms that preserve this gauge. Clearly these must be spatial diffeomorphisms, since a temporal diffeomorphism would move us away from $\delta\phi = 0$. Furthermore, for spatial diffeomorphisms that go to zero at spatial infinity, there is no additional freedom, but if we relax this requirement, we will find a set of nontrivial transformations.

We want to identify transformation rules for $\zeta,\gamma$ such that they can absorb the effect of a large diffeomorphism
\be
\delta\left(e^{2\zeta}\left(e^\gamma\right)_{ij}\right) = \pounds_\xi\left(e^{2\zeta}\left(e^\gamma\right)_{ij}\right).
\label{eq:gaugetrans}
\ee
We can formally solve this equation order by order in $\gamma$
\begin{eqnarray}
\nonumber
\xi_i &=& \xi_i^{(0)}+\xi_i^{(1)}+\ldots\\
\nonumber
\delta \zeta &=& \delta\zeta^{(0)}+\delta\zeta^{(1)}+\ldots\\
\delta\gamma_{ij} &=& \delta \gamma_{ij}^{(0)}+\delta \gamma_{ij}^{(1)}+\ldots 
\label{eq:gammaexpan}
\end{eqnarray}
However, the terms in the above that depend explicitly on $\gamma$ ({\it i.e.}, the first- and higher-order terms) will vanish because the $\gamma_{ij}$'s appearing in these terms are required to go to zero at infinity, and thus vanish in $(2+1)$-dimensions. The lowest order pieces (independent of the tensors) do not need to vanish, allowing for the possibility that a non-linear tensor profile is generated.

We therefore only need to work to lowest order in the tensors, henceforth we will drop the superscripts with this understanding. Equation~\eqref{eq:gaugetrans} leads to
\be
2\delta\zeta g_{ij}+\delta \gamma_{ij} = 2\xi_k\partial^k\zeta g_{ij}+\partial_i \xi_j+\partial_j\xi_i.
\label{eq:zetagammatransformations}
\ee
The nonlinear shift of the tensor modes must be traceless, so we can trace over both sides to obtain the transformation rule for $\zeta$,
\be
\delta\zeta = \frac{1}{2}\partial_i\xi^i+\xi^i\partial_i\zeta\,.
\ee
Substituting back into~\eqref{eq:zetagammatransformations}, we find the nonlinear shift of the tensor modes:
\be
\delta\gamma_{ij} = \partial_i\xi_j+\partial_j\xi_i-\partial_k\xi^k g_{ij}\,.
\ee
We must also impose the condition that $\delta\gamma_{ij}$ be transverse, which constrains the gauge parameter to be harmonic
\be
\partial\bar\partial\xi^i = 0\,.
\label{eq:harmoniccond}
\ee
This implies that we can write the components of $\xi^i$ as the sum of (possibly time-dependent) holomorphic and anti-holomorphic functions:
\be
\xi^i = f^i(z,\eta)+\bar g^i(\bar z, \eta)\,.
\label{eq:harmonicfns}
\ee

Though the diffeomorphisms~\eqref{eq:harmonicfns} preserve $\zeta$-gauge, in general they will not preserve the spatial initial value constraints of Einstein gravity. 
If we want the transformations to represent physical symmetries, we must check that they solve the equations of motion at finite momentum $k$ \cite{Weinberg:2003sw,Creminelli:2012ed,Hinterbichler:2012nm}. That is, we want to check that the profiles induced by the symmetries above arise as the $k\to0$ limit of physical modes which solve the constraint equations. This will constrain the time dependence of $\xi_i$.

The momentum and Hamiltonian constraints of General Relativity plus a scalar field can be solved at first order in the perturbations for the lapse function and the shift vector (see Appendix~\ref{Ddimzeta})
\be
N^{(1)} = \frac{\zeta'}{aH}\,;~~~~~~~~~~~~N_i^{(1)} = -\frac{1}{H}\partial_i\zeta +\frac{a\epsilon}{4c_s^2}\frac{\partial_i\zeta'}{\partial\bar\partial}\,.
\label{linearlapseandshift}
\ee
For the transformations~\eqref{eq:virzeta1} to be so-called adiabatic modes, they must preserve \eqref{linearlapseandshift}. 

The solutions to the constraints~\eqref{linearlapseandshift} are linear in $\zeta$, so we must only consider the non-linear field transformations. 
Under a spatial diffeomorphism, the fields transform as~\cite{Hinterbichler:2012nm}
\begin{align}
\nonumber
\delta\zeta &= \frac{1}{2}\partial_i\xi^i = \frac{1}{2}\left(\partial\xi^z+\bar\partial\xi^{\bar z}\right)\,,\\
\delta N^i &= \frac{1}{a}\xi'{}^i\, ,\\\nonumber
\delta N &= 0\,.
\end{align}
Inserting the lapse and shift transformation rules into~\eqref{linearlapseandshift} implies the following two equations
\begin{align}
\label{lapseconstraintxi}
&\frac{\rd}{\rd\eta}\left(\partial\xi^z+\bar\partial\xi^{\bar z}\right) = 0~,\\\label{shiftconstraintxi}
&\xi'{}^i = -\frac{a}{2H}g^{ik}\partial_k\left(\partial\xi^z+\bar\partial\xi^{\bar z}\right)\,,
\end{align}
along with the constraint that $\xi^i$ is harmonic. Here $g_{ij}$ is the spatial metric corresponding to the line element~\eqref{eq:dslineelement}.

In order to solve these equations, we first decompose the diffeomorphism parameter as
\be
\xi^i = \xi^i_0+\xi^i_T\,~~~~~{\rm where}~~~\partial_i\xi^i_T = \partial\xi^z_T+\bar\partial\xi^{\bar z}_T = 0\,.
\ee
This split is not unique, but can always be done. Equation~\eqref{lapseconstraintxi} implies that $\xi_0^i$ is time-independent, so the time dependence of $\xi$ is purely in the transverse part. Inserting this decomposition into~\eqref{shiftconstraintxi} yields the following equations for the time-dependent pieces:
\begin{align}
\xi_T'{}^z &= -\frac{1}{aH}\bar\partial^2\xi_0^{\bar z}\,,\\
\xi_T'{}^{\bar z} &= -\frac{1}{aH}\partial^2\xi_0^{z}\,.
\end{align}
Since $\xi_0^z$ and $\xi_0^{\bar z}$ are independent, let's first set  $\xi_0^{\bar z} = 0$ and consider $\xi_0^z$ which we can write as
\be
\xi_0^z = f^z_0(z)+\bar g^z_0(\bar z)\,;
\ee
For the time being, we will set $\bar g^z_0(\bar z) = 0$, but will return to this point shortly. 
Our choice of $f^z_0(z)$ fixes the time dependence of $\xi_T'{}^{\bar z}$ to be
\be
\xi_T'{}^{\bar z} = f'^{\bar z}(z,\eta)+\bar g'^{\bar z}(\bar z, \eta) = -\frac{1}{aH}\partial^2 f^z_0(z)\,.
\label{eq:barztimedepeq}
\ee
The right hand side of~\eqref{eq:barztimedepeq} is a function only of $z$, which implies that $\bar g'^{\bar z}(\bar z, \eta) = 0$, thus we are left with
\be
 f'^{\bar z}(z,\eta) = -\frac{1}{aH}\partial^2 f^z_0(z)\,.
\ee
We can now expand $f_0^z$ in a Laurent series $f_0^z(z)=-\sum_n a_nz^{n+1}$. 
Inserting this into~\eqref{shiftconstraintxi} yields the equations for each mode $n$
\be
\xi'{}^z_T = 0\,;~~~~~~~~~\xi'{}^{\bar z}_T = \frac{n(n+1)}{aH}z^{n-1}\,.
\ee
These equations imply that the $z$ component of $\xi$ remains time-independent, while the $\bar z$ component acquires a time-dependent contribution:
\be
\xi^z(z,\eta) = -z^{n+1}\,,~~~~~~~~~\xi^{\bar z}(z,\eta) = -n(n+1)z^{n-1}\int_{\eta}^0\frac{\rd \eta'}{aH}\,.
\label{eq:timedepxibar}
\ee

We therefore see that we are required to add a time-dependent $\xi^{\bar z}$ component in order for the $\xi^z$ transformation (which generates a conformal transformation of a spatial slice) to represent an adiabatic mode. 
Note that this time-dependence precisely matches that of the asymptotic symmetries when specialized to de Sitter backgrounds. Indeed, taking the de Sitter limit of~\eqref{eq:timedepxibar}, we can perform the integral, where we find $\xi^{\bar z}_T(z,\eta) = -\frac{\eta^2n(n+1)}{2}z^{n-1}$,
which agrees exactly with~\eqref{eq:Lnvectors}. On the other hand, at this order the adiabatic mode construction does not reproduce the temporal component of~\eqref{eq:Lnvectors}. This can be understood because $\zeta$ becomes time-independent in the $k\to0$ limit.

We could have considered instead a nonzero $\xi_0^{\bar z}$, which would lead to an independent set of anti-holomorphic adiabatic modes. These result in a time-dependent $\xi^z$ component, whose form can also be obtained simply by conjugating the holomorphic modes.

Finally, we would like to return to the case where we take $\xi_0^z$ to be purely a function of $\bar z$, and $\xi_0^{\bar z}$ to be purely a function of $z$:
\be
\xi_0^z = \bar g^z_0(\bar z)\,,~~~~~~~~~~~~\xi_0^{\bar z} = f^{\bar z}_0(z)\,.
\ee
This transformation is both time-independent and transverse, meaning that it does not acquire a time-dependent contribution from \eqref{shiftconstraintxi}. These transformations are the analogue of the ``tensor symmetries" of~\cite{Hinterbichler:2013dpa}. Indeed, it is straightforward to see that these transformations induce purely a nonlinear shift of the tensor modes, corresponding to an anisotropic rescaling.

This construction demonstrates that (at least in the 3-dimensional case) asymptotic symmetries and (scalar) adiabatic modes are two sides of the same coin. Either can be derived from the other. A similar viewpoint was espoused for QED in~\cite{Mirbabayi:2016xvc}. It is somewhat interesting that in addition to the scalar adiabatic modes, which capture the asymptotic symmetries, the set of adiabatic modes also contains a set of transformations which induce purely anisotropic profiles. These transformations do not decay sufficiently quickly at infinity to fall within the class of asymptotic symmetries, but rather lead to an anisotropic boundary metric. This seems to be quite closely related to the fall-off conditions considered in~\cite{Compere:2013bya,Avery:2013dja,Apolo:2014tua}. It would be interesting to understand this connection further.

\section{Ward identities}
\label{sec:Ward}

In this Section we derive the Ward identities associated with the asymptotic symmetries (or, equivalently, the adiabatic modes). To simplify the discussion, we only give the detailed derivation for the holomorphic symmetries and state the results for anti-holomorphic symmetries. 
The combined transformation of $\zeta$ under the asymptotic symmetries is 
\begin{align}
\nonumber
\delta_n\zeta &= -\frac{1}{2}(n+1)z^n-z^{n+1}\partial\zeta-n(n+1)\int_{\eta}^0\frac{\rd \eta'}{aH}\,z^{n-1}\bar\partial\zeta\,;\\
\bar{\delta}_n\zeta &= -\frac{1}{2}(n+1)\bar{z}^n-\bar{z}^{n+1}\bar{\partial}\zeta-n(n+1)\int_{\eta}^0\frac{\rd \eta'}{aH}\,\bar{z}^{n-1}\partial\zeta\,.
\label{eq:zetasymmetries}
\end{align}
These consist of a nonlinear shift, a time-independent linear piece, and a time-dependent linear piece. Each of these symmetries implies a Ward identity for correlation functions. Since these symmetries are non-linearly realized, they lead to soft theorems relating correlation functions with a soft $\zeta$ insertion to lower-point functions without the soft mode.  

 We derive these identities following the operator formalism of~\cite{Hinterbichler:2013dpa,Berezhiani:2014tda}. The transformations~\eqref{eq:zetasymmetries} are generated by the charges
\be
Q_n = \frac{1}{2}\int\rd z\rd\bar z\left\{\Pi_\zeta, \delta_n\zeta\right\}\,,
\ee
where $\Pi_\zeta = \delta{\cal L}/{\delta\dot\zeta}$ is the canonical momentum conjugate to $\zeta$. Similarly for $\bar{Q}_n$. In the quantum theory, the $Q_n$'s generate field transformations in the usual way as $[Q_n,\zeta] = -i\delta_n\zeta$. Given that these charges correspond to nonlinearly realized symmetries, formally they diverge. We will regulate this divergence by viewing the charge as the zero momentum limit of the charge in Fourier space.\footnote{Our conventions for Fourier transforms in terms of complex coordinates are given in Appendix~\ref{app:complexcoords}.} Physically this leads to soft theorems for $\zeta$. The only part of the charge for which we will need an explicit expression is the piece ${\cal Q}_n$ that generates the nonlinear transformation of $\zeta$, which is given by
\be
{\cal Q}_n = \lim_{q,\bar q\to 0} -(-2i)^n\frac{(n+1)}{2} \frac{\partial^n}{\partial \bar q^n}\tilde \Pi_\zeta(q,\bar q)\,,
\ee
where $\tilde \Pi_\zeta$ is the Fourier transform of the canonical momentum $\Pi$. This piece of the charge, ${\cal Q}_n$, is clearly time-independent. (The time-dependent part of the symmetry is linear in $\zeta$.)

We now wish to compute the in-in matrix element
\be
\langle\Omega\rvert [Q_n, {\cal O}]\lvert\Omega\rangle = -i\langle\Omega\rvert \delta_n{\cal O}\lvert\Omega\rangle\,,
\label{eq:wardiden}
\ee
where the operator ${\cal O}$ is an arbitrary product of $N$ $\zeta$'s with momenta $k_a^i$, $a = 1,\ldots,N$. To compute the left-hand side of~\eqref{eq:wardiden}, we require the action of the charge on the interacting vacuum\footnote{In deriving this, we have glossed over many technical subtleties (see~\cite{Hinterbichler:2013dpa,Berezhiani:2014tda} for a full discussion). The salient points are that we can relate the interacting vacuum to the free vacuum via time evolution: $\lvert\Omega\rangle = U^\dagger(-\infty,\eta)U_0(-\infty,\eta)\lvert0\rangle$, with $U$ the full time evolution operator, and $U_0$ its free-theory counterpart. We may then use the charge ${\cal Q}_n$ to evaluate the matrix element, as it is a symmetry of the free theory. Specifically $Q_n\lvert\Omega\rangle = Q_n U^\dagger(-\infty,\eta)U_0(-\infty,\eta)\lvert0\rangle = U^\dagger(-\infty,\eta)U_0(-\infty,\eta){\cal Q}_n\lvert0\rangle$. The action of ${\cal Q}_n$ on the free vacuum can then by computed from the knowledge that the Bunch--Davies vacuum wavefunctional is Gaussian. Finally, we use the fact that $\zeta$ is time-independent at long-wavelengths to write the free field in terms of the interacting operator $\zeta$, yielding~\eqref{eq:wardiden}.
}
\be
Q_n\lvert\Omega\rangle =  \lim_{q,\bar q\to 0}(-2i)^{n+1}\frac{(n+1)}{4} \frac{\partial^n}{\partial \bar q^n}\left( \frac{\zeta_{q,\bar q}}{2P_\zeta(q,\bar q)}\right)\lvert\Omega\rangle\,.
\ee
Using this, we can obtain for the left-hand side of~\eqref{eq:wardiden}:
\be
\langle\Omega\rvert [Q_n, {\cal O}]\lvert\Omega\rangle =  \lim_{q,\bar q\to 0}-(-2i)^{n+1}\frac{(n+1)}{4} \frac{\partial^n}{\partial \bar q^n}\left(\frac{\langle\zeta_{q,\bar q}{\cal O}\rangle}{P_\zeta(q,\bar q)} 
\right)\,.
\ee

In comparison, the right hand side of~\eqref{eq:wardiden} is easier to compute. The variation $\delta_n{\cal O}$ has two components: the nonlinear shift, and the linear piece. The nonlinear piece contributes a disconnected correlation function~\cite{Hinterbichler:2013dpa}, so by focusing on connected correlators (denoted by $\langle\cdots\rangle_c$) we only have to account for the linear transformation:
\be
-i\langle\Omega\rvert \delta_n{\cal O}\lvert\Omega\rangle_c =\frac{(n+1)}{2}(-2i)^{n+1}\sum_{a=1}^N\left(\frac{\partial^n}{\partial {\bar k}_a^n}+\frac{1}{(n+1)}\bar k_a\frac{\partial^{n+1}}{\partial{\bar k_a}^{n+1}}-\frac{n}{4}\int_{\eta}^0\frac{\rd \eta'}{aH}k_a\frac{\partial^{n-1}}{\partial\bar k_a^{n-1}}\right)\langle{\cal O}\rangle_c\,.
\ee
Putting these two pieces together we obtain the Ward identities 
\be
\lim_{q,\bar q\to 0}\frac{\partial^n}{\partial \bar q^n}\left(\frac{1}{2P_\zeta(q,\bar q)}\langle\zeta_{q,\bar q}\,{\cal O}\rangle_c\right)
= -\sum_{a=1}^N\left(\frac{\partial^n}{\partial {\bar k}_a^n}+\frac{1}{(n+1)}\bar k_a\frac{\partial^{n+1}}{\partial{\bar k_a}^{n+1}}-\frac{n}{4}\int_{\eta}^0\frac{\rd \eta'}{aH}\,k_a\frac{\partial^{n-1}}{\partial\bar k_a^{n-1}}\right)\langle{\cal O}\rangle_c\,.
\label{eq:zetavirasoroidentities}
\ee

Typically the most useful form of these identities involves removing the momentum-conserving delta function, so that the differential operators act on the primed correlators. In this case, the only difference between the primed and unprimed identities is that we sum only over $N-1$ momenta on the right hand side of the identity, and enforce momentum conservation by setting $k_N = -\sum_{a=1}^{N-1} k_a$, and similar for $\bar k_N$. See Appendix~\ref{app:fouriersymms} for more details. Additionally, we will be interested in checking the Ward identities in the $\eta\to0$ limit. In this limit, the final time-dependent term in~\eqref{eq:zetavirasoroidentities} vanishes, and we are left with\footnote{Even away from the $\eta\to0$ limit, the presence of the tensor symmetries allows us to remove the time-dependent part of the Ward identity~\cite{Hinterbichler:2013dpa}.}
\be
\lim_{q,\bar q\to 0}\frac{\partial^n}{\partial \bar q^n}\left(\frac{1}{2P_\zeta(q,\bar q)}\langle\zeta_{q,\bar q}\,{\cal O}\rangle_c'\right)=-\sum_{a=1}^{N-1}\left(\frac{\partial^{n}}{\partial {\bar k}_a^{n}}+\frac{1}{(n+1)}\bar k_a\frac{\partial^{n+1}}{\partial {\bar k}_a^{n+1}}\right)\langle {\cal O}\rangle_c'\,.
\label{eq:primedWardIDs}
\ee
This is the form of the identity we will check for $n=2$ in the following Section. Similarly it is straightforward to show that the anti-holomorphic transformations generated by $\bar{\delta}_n$ give rise to the Ward identities: 
\be
\lim_{q,\bar q\to 0}\frac{\partial^n}{\partial q^n}\left(\frac{1}{2P_\zeta(q,\bar q)}\langle\zeta_{q,\bar q}\,{\cal O}\rangle'_c\right)=-\sum_{a=1}^{N-1}\left(\frac{\partial^{n}}{\partial k_a^{n}}+\frac{1}{(n+1)}k_a\frac{\partial^{n+1}}{\partial k_a^{n+1}}\right)\langle {\cal O}\rangle'_c\,.
\label{eq:antiholomoprimedWardIDs}
\ee

In order to understand the physical consequence of these Ward identities, it is useful to recall the translation back into $(x,y)$ coordinates~\eqref{eq:tracelesssymshift}. As we saw earlier, the $\delta_n$ and $\bar\delta_n$ symmetries can be combined into a shift
where the nonlinear part consists of a shift of the field by a traceless symmetric matrix of coordinates. The Fourier-space consequence of this symmetry is that the traceless part of the ${\cal O}(q^n)$ soft limit is constrained in terms of lower point functions. Explicitly, in Cartesian coordinates, the Ward identities are
\begin{align}
\lim_{\vec q\to0} P^{i_1\cdots i_n}\frac{\partial^n}{\partial q^{i_1}\cdots\partial q^{i_n}}\left(\frac{1}{2P_\zeta(q)}\langle \zeta_q{\cal O}\rangle_c'\right) = &-P^{i_1\cdots i_n}\sum_{a=1}^{N-1}\Bigg[\frac{\partial^{n}}{\partial k_a^{i_1}\cdots\partial k_a^{i_n}} \\
\nonumber
&+ \frac{k_a^j}{n+1}\left(\delta_{i_1}^m\delta_j^\ell-\frac{1}{2}\delta_{i_1j}\delta^{\ell m}\right)\frac{\partial^{n+1}}{\partial k_a^\ell\partial k_a^{m}\partial k_a^{i_2}\cdots\partial k_a^{i_n}}\Bigg]\langle{\cal O}\rangle_c' \,,
\end{align}
where $P^{i_1\cdots i_n}$ is a projector onto symmetric, traceless tensors.

Here we have focused on the soft theorems following from the scalar symmetries, but there should also be some consequences due to the tensor adiabatic modes uncovered in Sec.~\ref{sec:zetasymms}. Presumably the identities following from these symmetries will involve insertions of boundary gravitons in correlation functions. Understanding the details of this would be very interesting.

\section{A check of the Ward identities}
\label{sec:check}
In this Section we give an example of an explicit check of the Ward identities derived in the previous Section.  We consider the $n=2$ identity for the scalar 3-point correlation function in the context of the effective field theory of inflation, in the limit of small speed of sound $c_s$. This identity relates the $O(q^2)$ component of the squeezed 3-point function to the 2-point function.

\subsection{EFT of inflation in $D$-dimensions}
The object we are ultimately interested in is the squeezed limit of the 3-point function for $\zeta$ in $(2+1)$-dimensions. Correlation functions with more than 2 insertions of $\zeta$ are somewhat difficult to compute in 2+1 dimensions because the mode functions are given by Hankel functions of integer order. Therefore, to make the computation tractable, we compute the squeezed 3-point function in a variety of even dimensions and ``dimensionally continue" to $D=3$.

Thus we start with the EFT of inflation~\cite{Creminelli:2006xe,Cheung:2007st} in $D$ dimensions. In this picture, inflation is viewed as a process of spontaneous breaking of time diffeomorphism invariance. In order to construct the effective theory, we begin by writing down all operators consistent with $d$-dimensional time-dependent diffeomorphism invariance\footnote{In this expression, we have not written explicitly operators which involve the extrinsic curvature because we are only interested in checking the soft-$\zeta$ theorem in a specific case, so we do not need the most general effective theory.}
\be
S = \int\rd^d x\rd t\sqrt{-g}\left(\frac{M_{\rm Pl}^{(d-1)}}{2}R-c(t) g^{00}-\Lambda(t)+\frac{M_2(t)}{2}(\delta g^{00})^2+\frac{M_3(t)}{3!}(\delta g^{00})^3+\ldots\right)\,,
\label{eq:EFTofinfaction}
\ee
where $\delta g^{00} = g^{00} + 1$.
We work in ADM variables, where the metric has the decomposition
\be
\rd s^2 = -N^2 \rd t^2 + h_{ij}\left(\rd x^i+N^i\rd t\right)\left(\rd x^j+N^j\rd t\right)\,.
\ee
Variation of the action~\eqref{eq:EFTofinfaction} with respect to the lapse and shift yield the background equations,
\begin{align}
\frac{d(d-1)}{2}M_{\rm Pl}^{(d-1)}H^2 &= c(t)+\Lambda(t)\, ;\\
M_{\rm Pl}^{(d-1)}\dot H &= -\frac{2}{(d-1)}c(t)\,.
\end{align}
These allow us to solve for $\Lambda(t)$ and $c(t)$.

The broken time diffeomorphism invariance can be restored by means of the St\"uckelberg replacement, $t\mapsto t+\pi$, leading to the action
\begin{align}
\nonumber
S = \int\rd^d x\rd t\sqrt{-g}\bigg(\frac{M_{\rm Pl}^{(d-1)}}{2}R+\frac{(d-1)M_{\rm Pl}^{(d-1)}}{2} \dot H& (Q-1)-\frac{(d-1)M_{\rm Pl}^{(d-1)}}{2}\left(d H^2+\dot H\right)\\
&+\frac{M_2(t+\pi)}{2}Q^2+\frac{M_3(t+\pi)}{3!}Q^3+\ldots\bigg)\,,
\label{eq:stuckelEFTinf}
\end{align}
where
\be
Q = 1 - N^{-2}\left(1+\dot\pi-N^i\partial_i\pi\right)^2 +h^{ij}\partial_i\pi\partial_j\pi\,.
\ee
In principle this action may be used to compute correlation functions of the field $\pi$ to arbitrary order in slow-roll by solving the constraints enforced by the lapse and shift. 

We are interested only in checking the Ward identities of the previous Section, so we will not do a fully general computation. Instead we focus on the decoupling limit
\be
M_{\rm Pl}\to \infty~~~~~~~~~~\dot H\to 0~~~~~~~~~~M^{(d-1)}_{\rm Pl}\dot H = {\rm fixed}\,.
\ee
In this limit $Q$ greatly simplifies: $Q \mapsto 1-(1+\dot\pi)^2+\frac{1}{a^2}(\nabla\pi)^2$, and the scale factor is that of de Sitter. Inserting this expression into~\eqref{eq:stuckelEFTinf}, we obtain the decoupling limit action for the Nambu--Goldstone field $\pi$
\begin{align}
\nonumber
S = \int\rd^d x\rd t\,a^d\bigg[\frac{(d-1)M_{\rm Pl}^{(d-1)}}{2}\dot H \left(-(1+\dot\pi)^2+\frac{1}{a^2}(\nabla\pi)^2\right)&+\frac{M_2}{2}\left(1-(1+\dot\pi)^2+\frac{1}{a^2}(\nabla\pi)^2\right)^2\\
&\!\!\!\!\!\!\!\!\!\!\!\!\!\!\!\!\!\!\!\!\!\!\!\!\!\!\!\!+\frac{M_3}{3!}\left(1-(1+\dot\pi)^2+\frac{1}{a^2}(\nabla\pi)^2\right)^3+\ldots\bigg]\,.
\end{align}
Our interest is in computing the 3-point function for $\pi$, so we expand this action to cubic order in $\pi$. Defining the sound speed of perturbations by
\be
c_s^{-2} \equiv 1 - \frac{4M_2}{(d-1)M^{(d-1)}_{\rm Pl}\dot H}\,,
\ee
the cubic action takes the following form
\be
\nonumber
S = -\frac{(d-1)M^{(d-1)}_{\rm Pl}\dot H}{2c_s^2}\int\rd^d x\rd t\,a^d\left(\dot \pi^2 - \frac{c_s^2}{a^2}(\nabla\pi)^2+(1-c_s^2)\left[\left(1-\frac{2M_3}{3M_2}\right)\dot \pi^3 - \frac{1}{a^2}\dot \pi(\nabla\pi)^2\right]\right)\,.
\ee
The soft theorems that we have derived should hold for any choice of the parameters $M_2, M_3$ in this action. Therefore, for simplicity we set $M_3 = 3M_2/2$, so that the $\dot\pi^3$ term drops out of the cubic action. We are left then with a single vertex $\sim \dot\pi(\nabla\pi)^2$:
\be
S = -\frac{(d-1)M^{(d-1)}_{\rm Pl}\dot H}{2c_s^2}\int\rd^d x\rd t\,a^d\left(\dot \pi^2 - \frac{c_s^2}{a^2}(\nabla\pi)^2+(c_s^2-1) \frac{1}{a^2}\dot \pi(\nabla\pi)^2\right)\,.
\label{eq:3ptaction}
\ee

\subsection{Mode functions and power spectrum}
We first derive the mode functions and power spectrum for the field $\pi$. 
In terms of the canonically-normalized field
\be
v_k = \left(\frac{(d-1)M^{(d-1)}_{\rm Pl} H^2\epsilon}{c_s^2}\right)^{1/2}a^\frac{d-1}{2}\pi_k\,,
\label{eq:cannormpi}
\ee
the quadratic equation of motion is
\be
v_k''+\left(c_s^2k^2 - \frac{1}{\eta^2}\left[\left(\frac{d}{2}\right)^2-\frac{1}{4}\right]\right)v_k = 0\,.
\ee
With the adiabatic vacuum initial conditions $\lim_{k\to\infty} v_k \simeq \frac{e^{-ic_sk\eta}}{\sqrt{2 c_s k}}$, the solution is a Hankel function of order $d/2$:

\be
v_k(\eta) = \frac{\pi^{1/2}e^{\frac{-(d+1) i\pi}{4}}}{2}(-\eta)^{1/2} H_{d/2}^{(1)}(-c_s k \eta)\,.
\ee
At long wavelengths ($k\eta\to0$), the power spectrum for the curvature perturbation, $\zeta = -H\pi$, is then given by
\be
\langle\zeta_k\zeta_k\rangle' = P_\zeta(k) = \frac{2^{d-2}H^{d-1}\Gamma\left[\frac{d}{2}\right]^2}{(d-1)\pi M_{\rm Pl}^{d-1}c_s^{d-2}\epsilon}\frac{1}{k^d}\,.
\ee

\subsection{Squeezed limit of the 3-point function}
In order to check the identities~\eqref{eq:primedWardIDs}, we need the squeezed limit of the 3-point function.
It ends up being most convenient to compute the squeezed limit directly, rather than computing the full answer. The interaction Hamiltonian derived from the action~\eqref{eq:3ptaction} is given by
\be
H_{\rm int.} = -\frac{(d-1)M_{\rm Pl}^{d-1}H^2\epsilon}{2}\left(1-c_s^{-2}\right)\int\rd^3x~ a^{d-2}\pi'(\nabla\pi)^2\,.
\ee
Following standard techniques, we arrive at the following integral expression for the $\langle\pi^3\rangle$ correlator
\begin{align}
 \langle\pi_{k_1}\pi_{k_2}\pi_{k_3}\rangle'  &=(d-1)M_{\rm Pl}^{d-1}H^2\epsilon (\vec k_2\cdot\vec k_3){\rm Im}\left\{ \pi^*_{k_1}(\eta)\pi^*_{k_2}(\eta)\pi^*_{k_3}(\eta)\int_{\infty^+}^0\rd\tilde\eta \frac{1}{(-H\tilde\eta)^{d-2}}\pi'_{k_1}(\tilde\eta)\pi_{k_2}(\tilde\eta)\pi_{k_3}(\tilde\eta)
\right\}\nn\\ 
&~~~~~~+5~{\rm perms.} \,
\label{eq:3ptintegral}
\end{align}
The squeezed limit may be taken directly in this integral. At leading order in $\vec k_1\to0$, this corresponds to setting $\vec k_2\cdot\vec k_3 = -k_2^2 - \vec k_1\cdot \vec k_2$ and $k_3=k_2+\frac{\vec k_1\cdot\vec k_2}{k_2}$.
When $d$ is even, this expression involves an integral over Hankel functions of integer order. In particular this includes the case of interest, $d=2$. Such expressions are difficult to integrate analytically. Therefore, we compute the squeezed limit in a somewhat indirect manner: we compute the squeezed 3-point function in a variety of cases where $d$ is odd in order to extract the expression in general dimension.\footnote{We of course have to assume that this expression continues to be correct in $d=2$. It would be preferable to compute the bispectrum directly in $d=2$.}
After converting to $\zeta$, we find the squeezed limit is given by (to leading nontrivial order)
\begin{align}
\label{eq:ddimsqueezed}
\lim_{k_1\to0}\langle\zeta_{k_1}\zeta_{k_2}\zeta_{k_3}\rangle' =  \frac{(D-1)}{3}P_\zeta(k_1)P_\zeta(k_2) (1-c_s^{-2}) \left((D-2)\frac{k_1^2}{k_2^2}- \frac{(D-3)(D+1)}{4}\frac{(\vec k_1\cdot\vec k_2)^2}{k_2^4}\right)\,.
\end{align}
Note that in $D=3$, the $(\vec k_1\cdot \vec k_2)^2$ piece vanishes. We will see that this is precisely what is required to satisfy the $n=2$ Ward identity. 

\subsection{Checking the $n=2$ identity}
Now that we have the squeezed limit to order $q^2$, we can use this to check some of the identities~\eqref{eq:primedWardIDs}. Rewriting the squeezed limit~\eqref{eq:ddimsqueezed} in $D=3$ and in $k,\bar k$ variables, we have
\be
\label{eq:d3squeezedexp}
\lim_{q,\bar q\to 0}\frac{1}{2P_\zeta(q,\bar q)}\langle\zeta_{q,\bar q}\zeta_{k_2,\bar k_2}\zeta_{k_3,\bar k_3}\rangle' = \frac{(1-c_s^{-2})P_\zeta(k_2,\bar k_2)}{3} \frac{q \bar q }{k_2 \bar k_2}+\ldots
\ee
Meanwhile, the power spectrum for $\zeta$ in $D=3$ is given by
\be
\langle\zeta_{k,\bar k}\zeta_{k,\bar k}\rangle' = \frac{H}{\pi\epsilon M_{\rm Pl}}\frac{1}{k\bar k}\,.
\ee
As a first check, we can see that the squeezed limit~\eqref{eq:d3squeezedexp} vanishes up to ${\cal O}(q^2)$. This implies that the 2-point function is annihilated by $\delta_0$ and $\delta_1$ (and their conjugates). Using the form~\eqref{eq:primedWardIDs} it is straightforward to check that this is the case. At ${\cal O}(q^2)$, the $n=2$ identity reads
\be
\label{eq:n2identity3pt}
\lim_{q,\bar q\to 0}\frac{\partial^2}{\partial \bar q^2}\left(\frac{1}{2P_\zeta(q,\bar q)}\langle\zeta_{q,\bar q}\zeta_{k_2,\bar k_2}\zeta_{k_3,\bar k_3}\rangle\right)
= -\left(\frac{\partial^2}{\partial {\bar k}^2}+\frac{1}{3}\bar k\frac{\partial^{3}}{\partial{\bar k}^{3}}\right)\langle\zeta_{k,\bar k}\zeta_{k,\bar k}\rangle'\,.
\ee
From~\eqref{eq:d3squeezedexp} we see that the left hand side vanishes --- there is no $\bar q^2$ or $q^2$ piece. This implies that the right hand side should also vanish. It is straightforward to check that the operator appearing on the right hand side of~\eqref{eq:n2identity3pt} annihilates the two point function (as does the corresponding operator for $\bar\delta_2$), so the $n=2$ relation is satisfied.

\section{Conclusions}
\label{sec:concl}

We have investigated how the asymptotic symmetry group of three dimensional de Sitter space acts on correlation functions during inflation. The infinite number of asymptotic symmetry transformations are in one-to-one correspondence with the adiabatic modes which generate nontrivial backgrounds for the curvature perturbation, $\zeta$. These nonlinearly realized symmetries lead to soft theorems which partially restrict the form of correlation functions in the soft limit -- where one of the momenta is taken to be smaller than the others -- at each order in the soft momentum. We have provided an initial check of these identities, including the first novel identity not present in higher-dimensional versions of inflation.

Looking forward, there are a number of interesting directions to pursue. Our analysis reinforces the notion that there is a close relationship between the asymptotic symmetries of a space-time, cosmological adiabatic modes, and soft theorems. It appears that the asymptotic symmetries are precisely those transformations which lead to physical soft insertions of the metric -- precisely what adiabatic modes are. It has been shown that higher-dimensional inflation possesses an infinite number of adiabatic modes~\cite{Hinterbichler:2012nm}, with corresponding Ward identities~\cite{Hinterbichler:2013dpa}. It would be interesting to understand these symmetries more directly in the language of asymptotic symmetries (see~\cite{Ferreira:2016hee} for some work in this direction). Further, the non-linearly realized components of the four-dimensional symmetries are similar to the ones discussed in this note: they consist of traceless polynomials in the coordinates. It would be interesting to understand the relationship between these two sets of symmetries, perhaps through dimensional reduction.

Staying within the realm of three dimensional inflation there are also a number of intriguing directions to pursue. It has long been known that three-dimensional gravity possesses a formulation as a Chern--Simons theory with ${\rm SL}(2,{\mathbb R})\times{\rm SL}(2,{\mathbb R})$ gauge group \cite{Achucarro:1987vz,Witten:1988hc}. It would be very interesting to consider inflation from this perspective, as opposed to the metic formulation we have used here. It may also be interesting to consider generalizations of the identities we have considered, for example to multiple soft external legs, along the lines of~\cite{Joyce:2014aqa,Mirbabayi:2014zpa}. In~\cite{Kaplan:2014dia}, the EFT of holographic RG flows was considered, which is essentially inflation on its side; it would be interesting to understand whether the soft theorems for the dilaton have any important consequences in that setup.

 Additionally, we have focused primarily on the dynamics and Ward identities of the curvature perturbation itself; it should be possible to extend these arguments to include correlation functions of spectator fields during inflation. Since the vacuum is only invariant under the global SL$(2,{\mathbb C})$ subgroup of the Virasoro symmetries, we expect that correlation functions of spectator fields will only be annihilated by $\delta_{-1},\delta_0, \delta_1$ and their conjugates (indeed, this can be explicitly checked).  The higher Virasoro symmetries correspond to insertions of ``boundary gravitons,'' which correspond to insertions of moments of the stress tensor in the dual CFT, leading to a late-time identity of the form (see also~\cite{Fitzpatrick:2014vua})
\be
\langle L_n\phi(z_1)\phi(z_2)\cdots\phi(z_N)\rangle = -\sum_{a=1}^{N}\left(\frac{(n+1)\Delta_a}{2}z_a^{n}+z_a^{n+1}\right)\langle\phi(z_1)\phi(z_2)\cdots\phi(z_N)\rangle\,,
\label{eq:specIDs}
\ee
where $L_n$ is the operator which generates the vector field $\ell_n$ on the boundary, $z_a$ are points on the future boundary, and $\Delta$ is set by the mass of the field $\phi$ through $\Delta = 1-\sqrt{1-\frac{m_\phi^2}{H^2}}$. It would be interesting to derive the identity~\eqref{eq:specIDs} directly from the bulk physics, and check it on multi-field correlators. Similarly, it would be interesting to investigate the consequences of the purely tensor adiabatic modes and elucidate their relation to non-Brown--Henneaux boundary conditions.

Three-dimensional inflation provides a simplified arena in which to explore the relationship between adiabatic modes, asymptotic symmetries, and soft theorems in the context of cosmology. 
Aside from being of interest in their own right, we hope that the lessons abstracted from this situation may be usefully applied to gain a better understanding of these phenomena in higher dimensions.

\vspace{-.4cm}
\paragraph{Acknowledgements:} We thank Paolo Creminelli, Wayne Hu, Lam Hui, Emil Martinec, Rachel Rosen, Marko Simonovi\'c and Junpu Wang for helpful discussions.  We would like to thank the Sitka Sound Science Center for hospitality while some of this work was completed (A.J. \& K.H.). This work was supported in part by NASA ATP grant NNX16AB27G, National Science Foundation Grant No. PHYS-1066293 and the hospitality of the Aspen Center for Physics, and also by the Kavli Institute for Cosmological Physics at the University of Chicago through grant NSF PHY-1125897, an endowment from the Kavli Foundation and its founder Fred Kavli, and by the Robert R. McCormick Postdoctoral Fellowship (A.J.). J.K. is supported in part by NSF CAREER Award PHY-1145525.

\appendix

\section{Complex coordinates}
\label{app:complexcoords}

In two spatial dimensions, it is often useful to work in terms of complexified coordinates, 
\be z = x+iy\, , \ \ \ \bar z = x-iy\, .\ee
In these coordinates, the non-zero metric components are
 \be g_{z\bar z} = g_{\bar z z} =\frac{1}{2},\ \ \  g^{z\bar z} =g^{\bar zz}= 2\ ,\ee
 so there are sometimes hidden factors of 2 to account for.  A contravariant vector $v^i$ is split into complex components according to
\be
v^z = v^x+i v^y\,;~~~~~~~v^{\bar z} = v^x-i v^y\,;~~~~~~~v_{z} = \frac{1}{2}(v^x-i v^y)\,;~~~~~~~v_{\bar z} =  \frac{1}{2}(v^x+i v^y)\,.
\ee
For vectors, we will often define $ v \equiv v^z$ and $\bar v \equiv v^{\bar z}$. With this convention, the Cartesian dot product translated to complex coordinates is given by $\vec a\cdot\vec b = \frac{1}{2}(a\bar b+\bar a b)$. Derivative operators are
\be
\partial \equiv \partial_z = \frac{1}{2}(\partial_x-i\partial_y),~~~~~~~\bar\partial \equiv \partial_{\bar z}= \frac{1}{2}(\partial_x+i\partial_y)\,.
\ee
This definition implies that $\nabla^2 = \delta^{ij}\partial_i\partial_j = 4\partial\bar\partial$. For many applications we work in Fourier space. This is obtained by implementing the above coordinate change in the Cartesian Fourier transform:
\be
f(\vec x) = \int\frac{\rd^2 k}{(2\pi)^2} e^{-i\vec k\cdot\vec x}\tilde f (\vec k) \ \ \Leftrightarrow \ \  f(z,\bar z)= \frac{1}{2} \int\frac{\rd k\rd \bar k}{(2\pi)^2}e^{-\frac{i}{2}(k\bar z+\bar k z)}\tilde f (k,\bar k)\,.
\label{eq:fouriertrans}
\ee

\section{Constraint equations in $(d+1)$-dimensions}
\label{Ddimzeta}
In section~\ref{sec:zetasymms}, in order to construct the adiabatic modes corresponding to the $\zeta$ symmetries, we required the explicit solutions for the constraint equations implied by the lapse and shift variables in $(2+1)$ dimensions. In this Appendix, we derive the relevant constraint equations and their solutions. We work in general dimension ($d$ is the number of spatial dimensions) and specialize to the case of interest at the end.

Consider the Einstein--Hilbert action in $(d+1)$-dimensions coupled to a $P(X,\phi)$ theory, where $X\equiv -{1\over 2}(\partial\phi)^2$,
\be
S = \frac{1}{2}\int\rd^{d+1}x\sqrt{-g}\left(M_{\rm Pl}^{(d-1)}R+2P(X,\phi)\right)\,.
\ee
In terms of ADM variables, where the line element is given by
\be
\rd s^2 = -N^2 \rd t^2 + h_{ij}\left(\rd x^i+N^i\rd t\right)\left(\rd x^j+N^j\rd t\right)\,,
\ee
and using the decomposition of the Ricci scalar, $R = R^{(d)} +K^{ij}K_{ij} - K^2$,
where the extrinsic curvature is
\be
K_{ij} = {1\over N}E_{ij} = \frac{1}{2N}\left(\dot h_{ij}-\nabla_iN_j-\nabla_j N_i\right)\,,
\ee
the action takes the following form
\be
S = \frac{M_{\rm Pl}^{(d-1)}}{2}\int\rd^{d+1}x\sqrt{h}N\left[R^{(d)}+N^{-2}\left(E^i_jE^j_i-E^2\right)+2P\left(\frac{1}{2N^2}(\dot\phi - N^i\partial_i\phi)^2-\frac{1}{2}(\nabla\phi)^2, \phi\right)\right]\,.
\label{ddimehaction}
\ee
We are interested in perturbations about FRW backgrounds with homogeneous field profiles ($\phi = \phi(t)$). Furthermore we choose $\zeta$-gauge, defined by
\be
\delta\phi = 0~,~~~~~~~~~~~~h_{ij}=a^2 e^{2\zeta}(e^\gamma)_{ij}~;~~~~~~~~~~~~\gamma^i_i = \partial_i\gamma^{ij} = 0\,.
\label{eq:zetagauge}
\ee
We focus on the scalar sector, setting $\gamma_{ij} = 0$. Varying the action~\eqref{ddimehaction} with respect to $N$, we obtain the Hamiltonian constraint
\be
R^{(d)} - N^{-2}\left(E^i_jE^j_i-E^2\right)+2P-4XP_{,X} = 0~,
\ee
and varying with respect to the shift, we obtain the momentum constraint:
\be
\nabla_i \left[N^{-1}\left(E^i_j - E\delta^i_j\right)\right] = 0~.
\ee
We define the first order perturbations of the lapse and shift through
\be
N = 1+N^{(1)}\,;~~~~~~~~~~~N^i = N_i^T+\partial_i\psi\,.
\ee
Inserting the gauge choice~\eqref{eq:zetagauge}, the constraint equations at first order in the perturbations read
\begin{align}
\label{eq:momentumconst}
&(d-1)\partial_i\left(HN^{(1)}-\dot\zeta\right)-\frac{1}{2}a^{-2}\nabla^2N_i^T = 0\,;\\
\label{eq:hamiltonianconst}
&-2(d-1)a^{-2}\partial_i\left(\partial^i\zeta+H\partial^i\psi\right)+\frac{2(d-1)H^2\epsilon}{c_s^2}N^{(1)}-2d(d-1)H\left(HN^{(1)}-\dot\zeta\right) =0\,.
\end{align}
The momentum constraint~\eqref{eq:momentumconst} is solved by
\be
N^{(1)} = \frac{\dot\zeta}{H}~;~~~~~~~~~~N^T_i = 0~,
\ee
we can then plug this into the Hamiltonian constraint~\eqref{eq:hamiltonianconst} to obtain:
\be
a^{-2}\nabla^2\zeta+Ha^{-2}\nabla^2\psi-\frac{H\epsilon}{c_s^2}\dot\zeta = 0\,.
\ee
This can be formally inverted to obtain
\be
N_i^{(1)} = \partial_i\psi~;~~~~~~~~\psi =-\frac{\zeta}{H} +\frac{a^2\epsilon}{c_s^2}\frac{1}{\nabla^2}\dot\zeta\,.
\ee
Specializing to $d=2$ and writing in terms of conformal time then yields~\eqref{linearlapseandshift}.

\section{Symmetry transformations in Fourier space}
\label{app:fouriersymms}
In the text, we utilize the action of the asymptotic symmetries on fields in Fourier space. In this Appendix, we derive these expression for the symmetry variations. We only need consider the linear transformation of the fields under the asymptotic symmetries:
\be
\delta_n\phi = -\left(\frac{(n+1)\Delta}{2}z^{n}+z^{n+1}\partial\right)\phi\,,
\ee
where the linear transformation of $\zeta$ corresponds to a $\Delta = 0$ field. We can then derive the transformation rules by going to Fourier space using~\eqref{eq:fouriertrans}
\begin{align}
\nonumber
\delta_n\phi &= -\left(\frac{(n+1)\Delta}{2}z^{n}+z^{n+1}\partial\right)\frac{1}{2}\int\frac{\rd k\rd\bar k}{(2\pi)^2}e^{-\frac{i}{2}(k\bar z+\bar k z)}\phi_{k,\bar k}\\
 &= -\frac{(2i)^n}{2}\int\frac{\rd k\rd\bar k}{(2\pi)^2}\,\phi_{k,\bar k}\left(\frac{(n+1)\Delta}{2}\frac{\partial^n}{\partial {\bar k}^n}+\bar k\frac{\partial^{n+1}}{\partial {\bar k}^{n+1}}\right)e^{-\frac{i}{2}(k\bar z+\bar k z)} \\\nonumber
 &=-\frac{(n+1)(-2i)^n}{2}\int\frac{\rd k\rd\bar k}{(2\pi)^2}e^{-\frac{i}{2}(k\bar z+\bar k z)}\left(\left(\frac{\Delta}{2}-1\right)\frac{\partial^{n}}{\partial {\bar k}^{n}}-\frac{1}{(n+1)}\bar k\frac{\partial^{n+1}}{\partial {\bar k}^{n+1}}\right)\phi_{k,\bar k}\,,
\end{align}
which then implies the Fourier space transformation rule
\be
\delta_n\phi_{k,\bar k} = (n+1)(-2i)^n\left(\left(1-\frac{\Delta}{2}\right)\frac{\partial^{n}}{\partial {\bar k}^{n}}+\frac{1}{(n+1)}\bar k\frac{\partial^{n+1}}{\partial {\bar k}^{n+1}}\right)\phi_{k,\bar k}\,.
\label{eq:fspacevirasoros}
\ee
%
\paragraph{Action on correlators:} Typically we will be interested in the variation of correlation functions under the symmetries~\eqref{eq:fspacevirasoros}. This bring with it an additional subtlety, a general correlation function is of the form\footnote{Note that $\delta(\Sigma_a\vec k_a)=4\delta(\Sigma_a k_a+\bar k_a)\delta(\Sigma_a k_a-\bar k_a)=4\delta(\Sigma_a k_a)\delta(\Sigma_a \bar k_a)$.}
\be
\langle {\cal O}_1(k_1,\bar k_1)\cdots{\cal O}_N(k_N\bar k_N)\rangle = 4(2\pi)^2\delta(\Sigma_a k_a)\delta(\Sigma_a \bar k_a)\langle {\cal O}_1(k_1,\bar k_1)\cdots{\cal O}_N(k_N\bar k_N)\rangle'\,,
\ee
and generally the operators~\eqref{eq:fspacevirasoros} will also hit the delta function outside the correlator which enforces momentum conservation. It is therefore desirable to write identities directly in terms of the primed correlation functions. The problem of commuting differential operators of this type through delta functions has been considered very carefully in~\cite{Berezhiani:2014tda} and we can apply their results more or less directly to the case of interest. The essential point is that in the final Ward identities, the contributions from operators hitting the delta function are always proportional to lower-order identities, and therefore vanish.

Therefore, the effective result of commuting through the delta function is that the operators act only on $N-1$ of the momenta in the correlation function (as the delta function restricts $k_N = -\Sigma_a k_a$) so that the operators take the form:
\be
\delta_n\langle {\cal O}_1\cdots{\cal O}_N\rangle' = (n+1)(-2i)^n\left(-\delta_{n0}\frac{\Delta_N}{2}+\sum_{a=1}^{N-1}\left[\left(1-\frac{\Delta_a}{2}\right)\frac{\partial^{n}}{\partial {\bar k}^{n}}+\frac{1}{(n+1)}\bar k\frac{\partial^{n+1}}{\partial {\bar k}^{n+1}}\right]\right)\langle {\cal O}_1\cdots{\cal O}_N\rangle'\,.
\ee
The only subtlety is that in the $n=0$ case we must subtract the conformal weights of ${\it all}$ the operators. Beyond that, the identities for full correlators can be written for primed correlators by specializing to the locus in momentum space where the delta function is satisfied. The other operators which arise in the text, $\frac{\partial^n}{\partial q^n}$ and $k\frac{\partial^{n-1}}{\partial \bar k^{n-1}}$ can be commuted through the delta function in the same way.

\renewcommand{\em}{}
\bibliographystyle{utphys}
\addcontentsline{toc}{section}{References}
\bibliography{3dinflationdraft17}

\providecommand{\href}[2]{#2}\begingroup\raggedright\begin{thebibliography}{10}

\bibitem{abbott1884flatland}
E.~Abbott, {\em Flatland: A Romance of Many Dimensions}.
\newblock Seeley, 1884.
\newblock \url{https://books.google.com/books?id=139bAAAAQAAJ}.

\bibitem{Deser:1983tn}
S.~Deser, R.~Jackiw, and G.~'t~Hooft, ``{Three-Dimensional Einstein Gravity:
  Dynamics of Flat Space},''
\href{http://dx.doi.org/10.1016/0003-4916(84)90085-X}{{\em Annals Phys.} {\bf
  152} (1984)  220}.

\bibitem{Banados:1992wn}
M.~Banados, C.~Teitelboim, and J.~Zanelli, ``{The Black hole in
  three-dimensional space-time},''
  \href{http://dx.doi.org/10.1103/PhysRevLett.69.1849}{{\em Phys. Rev. Lett.}
  {\bf 69} (1992)  1849--1851},
\href{http://arxiv.org/abs/hep-th/9204099}{{\tt arXiv:hep-th/9204099
  [hep-th]}}.

\bibitem{Banados:1992gq}
M.~Banados, M.~Henneaux, C.~Teitelboim, and J.~Zanelli, ``{Geometry of the
  (2+1) black hole},'' \href{http://dx.doi.org/10.1103/PhysRevD.48.1506,
  10.1103/PhysRevD.88.069902}{{\em Phys. Rev.} {\bf D48} (1993)  1506--1525},
  \href{http://arxiv.org/abs/gr-qc/9302012}{{\tt arXiv:gr-qc/9302012 [gr-qc]}}.
[Erratum: Phys. Rev.D88,069902(2013)].

\bibitem{Carlip:1995zj}
S.~Carlip, ``{Lectures on (2+1) dimensional gravity},'' {\em J. Korean Phys.
  Soc.} {\bf 28} (1995)  S447--S467,
\href{http://arxiv.org/abs/gr-qc/9503024}{{\tt arXiv:gr-qc/9503024 [gr-qc]}}.

\bibitem{Martinec:2014uva}
E.~J. Martinec and W.~E. Moore, ``{Modeling Quantum Gravity Effects in
  Inflation},'' \href{http://dx.doi.org/10.1007/JHEP07(2014)053}{{\em JHEP}
  {\bf 07} (2014)  053},
\href{http://arxiv.org/abs/1401.7681}{{\tt arXiv:1401.7681 [hep-th]}}.

\bibitem{Moore:2014sia}
W.~E. Moore, ``{Primordial fluctuations in extended Liouville theory},''
  \href{http://dx.doi.org/10.1007/JHEP03(2015)001}{{\em JHEP} {\bf 03} (2015)
  001},
\href{http://arxiv.org/abs/1411.2612}{{\tt arXiv:1411.2612 [hep-th]}}.

\bibitem{Banks:1984np}
T.~Banks, W.~Fischler, and L.~Susskind, ``{Quantum Cosmology in
  (2+1)-dimensions and (3+1)-dimensions},''
\href{http://dx.doi.org/10.1016/0550-3213(85)90070-7}{{\em Nucl. Phys.} {\bf
  B262} (1985)  159--186}.

\bibitem{Samiullah:1991qv}
M.~Samiullah, O.~J.~P. Eboli, and S.-Y. Pi, ``{Quantum field theoretic analysis
  of inflation dynamics in a planar universe},''
\href{http://dx.doi.org/10.1103/PhysRevD.44.2335}{{\em Phys. Rev.} {\bf D44}
  (1991)  2335--2355}.

\bibitem{Kaplan:2014dia}
J.~Kaplan and J.~Wang, ``{An Effective Theory for Holographic RG Flows},''
  \href{http://dx.doi.org/10.1007/JHEP02(2015)056}{{\em JHEP} {\bf 02} (2015)
  056},
\href{http://arxiv.org/abs/1406.4152}{{\tt arXiv:1406.4152 [hep-th]}}.

\bibitem{Creminelli:2006xe}
P.~Creminelli, M.~A. Luty, A.~Nicolis, and L.~Senatore, ``{Starting the
  Universe: Stable Violation of the Null Energy Condition and Non-standard
  Cosmologies},'' \href{http://dx.doi.org/10.1088/1126-6708/2006/12/080}{{\em
  JHEP} {\bf 12} (2006)  080},
\href{http://arxiv.org/abs/hep-th/0606090}{{\tt arXiv:hep-th/0606090
  [hep-th]}}.

\bibitem{Cheung:2007st}
C.~Cheung, P.~Creminelli, A.~L. Fitzpatrick, J.~Kaplan, and L.~Senatore, ``{The
  Effective Field Theory of Inflation},''
  \href{http://dx.doi.org/10.1088/1126-6708/2008/03/014}{{\em JHEP} {\bf 03}
  (2008)  014},
\href{http://arxiv.org/abs/0709.0293}{{\tt arXiv:0709.0293 [hep-th]}}.

\bibitem{Creminelli:2012ed}
P.~Creminelli, J.~Norena, and M.~Simonovi\'c, ``{Conformal consistency
  relations for single-field inflation},''
  \href{http://dx.doi.org/10.1088/1475-7516/2012/07/052}{{\em JCAP} {\bf 1207}
  (2012)  052},
\href{http://arxiv.org/abs/1203.4595}{{\tt arXiv:1203.4595 [hep-th]}}.

\bibitem{Hinterbichler:2012nm}
K.~Hinterbichler, L.~Hui, and J.~Khoury, ``{Conformal Symmetries of Adiabatic
  Modes in Cosmology},''
  \href{http://dx.doi.org/10.1088/1475-7516/2012/08/017}{{\em JCAP} {\bf 1208}
  (2012)  017},
\href{http://arxiv.org/abs/1203.6351}{{\tt arXiv:1203.6351 [hep-th]}}.

\bibitem{Assassi:2012zq}
V.~Assassi, D.~Baumann, and D.~Green, ``{On Soft Limits of Inflationary
  Correlation Functions},''
  \href{http://dx.doi.org/10.1088/1475-7516/2012/11/047}{{\em JCAP} {\bf 1211}
  (2012)  047},
\href{http://arxiv.org/abs/1204.4207}{{\tt arXiv:1204.4207 [hep-th]}}.

\bibitem{Maldacena:2002vr}
J.~M. Maldacena, ``{Non-Gaussian features of primordial fluctuations in single
  field inflationary models},''
  \href{http://dx.doi.org/10.1088/1126-6708/2003/05/013}{{\em JHEP} {\bf 0305}
  (2003)  013},
\href{http://arxiv.org/abs/astro-ph/0210603}{{\tt arXiv:astro-ph/0210603
  [astro-ph]}}.

\bibitem{Creminelli:2004yq}
P.~Creminelli and M.~Zaldarriaga, ``{Single field consistency relation for the
  3-point function},''
  \href{http://dx.doi.org/10.1088/1475-7516/2004/10/006}{{\em JCAP} {\bf 0410}
  (2004)  006},
\href{http://arxiv.org/abs/astro-ph/0407059}{{\tt arXiv:astro-ph/0407059
  [astro-ph]}}.

\bibitem{Hinterbichler:2013dpa}
K.~Hinterbichler, L.~Hui, and J.~Khoury, ``{An Infinite Set of Ward Identities
  for Adiabatic Modes in Cosmology},''
  \href{http://dx.doi.org/10.1088/1475-7516/2014/01/039}{{\em JCAP} {\bf 1401}
  (2014)  039},
\href{http://arxiv.org/abs/1304.5527}{{\tt arXiv:1304.5527 [hep-th]}}.

\bibitem{Kehagias:2012pd}
A.~Kehagias and A.~Riotto, ``{Operator Product Expansion of Inflationary
  Correlators and Conformal Symmetry of de Sitter},''
  \href{http://dx.doi.org/10.1016/j.nuclphysb.2012.07.004}{{\em Nucl. Phys.}
  {\bf B864} (2012)  492--529},
\href{http://arxiv.org/abs/1205.1523}{{\tt arXiv:1205.1523 [hep-th]}}.

\bibitem{Goldberger:2013rsa}
W.~D. Goldberger, L.~Hui, and A.~Nicolis, ``{One-particle-irreducible
  consistency relations for cosmological perturbations},''
  \href{http://dx.doi.org/10.1103/PhysRevD.87.103520}{{\em Phys. Rev.} {\bf
  D87} (2013) no.~10, 103520},
\href{http://arxiv.org/abs/1303.1193}{{\tt arXiv:1303.1193 [hep-th]}}.

\bibitem{Adler:1964um}
S.~L. Adler, ``{Consistency conditions on the strong interactions implied by a
  partially conserved axial vector current},''
\href{http://dx.doi.org/10.1103/PhysRev.137.B1022}{{\em Phys. Rev.} {\bf 137}
  (1965)  B1022--B1033}.

\bibitem{Schalm:2012pi}
K.~Schalm, G.~Shiu, and T.~van~der Aalst, ``{Consistency condition for
  inflation from (broken) conformal symmetry},''
  \href{http://dx.doi.org/10.1088/1475-7516/2013/03/005}{{\em JCAP} {\bf 1303}
  (2013)  005},
\href{http://arxiv.org/abs/1211.2157}{{\tt arXiv:1211.2157 [hep-th]}}.

\bibitem{Mata:2012bx}
I.~Mata, S.~Raju, and S.~Trivedi, ``{CMB from CFT},''
  \href{http://dx.doi.org/10.1007/JHEP07(2013)015}{{\em JHEP} {\bf 07} (2013)
  015},
\href{http://arxiv.org/abs/1211.5482}{{\tt arXiv:1211.5482 [hep-th]}}.

\bibitem{McFadden:2014nta}
P.~McFadden, ``{Soft limits in holographic cosmology},''
  \href{http://dx.doi.org/10.1007/JHEP02(2015)053}{{\em JHEP} {\bf 02} (2015)
  053},
\href{http://arxiv.org/abs/1412.1874}{{\tt arXiv:1412.1874 [hep-th]}}.

\bibitem{Berezhiani:2013ewa}
L.~Berezhiani and J.~Khoury, ``{Slavnov-Taylor Identities for Primordial
  Perturbations},'' \href{http://dx.doi.org/10.1088/1475-7516/2014/02/003}{{\em
  JCAP} {\bf 1402} (2014)  003},
\href{http://arxiv.org/abs/1309.4461}{{\tt arXiv:1309.4461 [hep-th]}}.

\bibitem{Pimentel:2013gza}
G.~L. Pimentel, ``{Inflationary Consistency Conditions from a Wavefunctional
  Perspective},'' \href{http://dx.doi.org/10.1007/JHEP02(2014)124}{{\em JHEP}
  {\bf 02} (2014)  124},
\href{http://arxiv.org/abs/1309.1793}{{\tt arXiv:1309.1793 [hep-th]}}.

\bibitem{Collins:2014fwa}
H.~Collins, R.~Holman, and T.~Vardanyan, ``{A Cosmological Slavnov-Taylor
  Identity},'' \href{http://dx.doi.org/10.1088/1475-7516/2014/12/007}{{\em
  JCAP} {\bf 1412} (2014) no.~12, 007},
\href{http://arxiv.org/abs/1405.0017}{{\tt arXiv:1405.0017 [hep-th]}}.

\bibitem{Armendariz-Picon:2014xda}
C.~Armendariz-Picon, J.~T. Neelakanta, and R.~Penco, ``{General Covariance
  Constraints on Cosmological Correlators},''
  \href{http://dx.doi.org/10.1088/1475-7516/2015/01/035}{{\em JCAP} {\bf 1501}
  (2015) no.~01, 035},
\href{http://arxiv.org/abs/1411.0036}{{\tt arXiv:1411.0036 [hep-th]}}.

\bibitem{Berezhiani:2014kga}
L.~Berezhiani and J.~Khoury, ``{On the Initial State and Consistency
  Relations},'' \href{http://dx.doi.org/10.1088/1475-7516/2014/09/018}{{\em
  JCAP} {\bf 1409} (2014)  018},
\href{http://arxiv.org/abs/1406.2689}{{\tt arXiv:1406.2689 [hep-th]}}.

\bibitem{Berezhiani:2014tda}
L.~Berezhiani, J.~Khoury, and J.~Wang, ``{Non-Trivial Checks of Novel
  Consistency Relations},''
  \href{http://dx.doi.org/10.1088/1475-7516/2014/06/056}{{\em JCAP} {\bf 1406}
  (2014)  056},
\href{http://arxiv.org/abs/1401.7991}{{\tt arXiv:1401.7991 [hep-th]}}.

\bibitem{Ferreira:2016hee}
R.~Z. Ferreira, M.~Sandora, and M.~S. Sloth, ``{Asymptotic Symmetries in de
  Sitter and Inflationary Spacetimes},''
\href{http://arxiv.org/abs/1609.06318}{{\tt arXiv:1609.06318 [hep-th]}}.

\bibitem{Joyce:2014aqa}
A.~Joyce, J.~Khoury, and M.~Simonovi\'c, ``{Multiple Soft Limits of
  Cosmological Correlation Functions},''
  \href{http://dx.doi.org/10.1088/1475-7516/2015/01/012}{{\em JCAP} {\bf 1501}
  (2015) no.~01, 012},
\href{http://arxiv.org/abs/1409.6318}{{\tt arXiv:1409.6318 [hep-th]}}.

\bibitem{Mirbabayi:2014zpa}
M.~Mirbabayi and M.~Zaldarriaga, ``{Double Soft Limits of Cosmological
  Correlations},'' \href{http://dx.doi.org/10.1088/1475-7516/2015/03/025}{{\em
  JCAP} {\bf 1503} (2015) no.~03, 025},
\href{http://arxiv.org/abs/1409.6317}{{\tt arXiv:1409.6317 [hep-th]}}.

\bibitem{Kehagias:2013yd}
A.~Kehagias and A.~Riotto, ``{Symmetries and Consistency Relations in the Large
  Scale Structure of the Universe},''
  \href{http://dx.doi.org/10.1016/j.nuclphysb.2013.05.009}{{\em Nucl. Phys.}
  {\bf B873} (2013)  514--529},
\href{http://arxiv.org/abs/1302.0130}{{\tt arXiv:1302.0130 [astro-ph.CO]}}.

\bibitem{Peloso:2013zw}
M.~Peloso and M.~Pietroni, ``{Galilean invariance and the consistency relation
  for the nonlinear squeezed bispectrum of large scale structure},''
  \href{http://dx.doi.org/10.1088/1475-7516/2013/05/031}{{\em JCAP} {\bf 1305}
  (2013)  031},
\href{http://arxiv.org/abs/1302.0223}{{\tt arXiv:1302.0223 [astro-ph.CO]}}.

\bibitem{Creminelli:2013mca}
P.~Creminelli, J.~Noreña, M.~Simonovi\'c, and F.~Vernizzi, ``{Single-Field
  Consistency Relations of Large Scale Structure},''
  \href{http://dx.doi.org/10.1088/1475-7516/2013/12/025}{{\em JCAP} {\bf 1312}
  (2013)  025},
\href{http://arxiv.org/abs/1309.3557}{{\tt arXiv:1309.3557 [astro-ph.CO]}}.

\bibitem{Creminelli:2013poa}
P.~Creminelli, J.~Gleyzes, M.~Simonovi\'c, and F.~Vernizzi, ``{Single-Field
  Consistency Relations of Large Scale Structure. Part II: Resummation and
  Redshift Space},''
  \href{http://dx.doi.org/10.1088/1475-7516/2014/02/051}{{\em JCAP} {\bf 1402}
  (2014)  051},
\href{http://arxiv.org/abs/1311.0290}{{\tt arXiv:1311.0290 [astro-ph.CO]}}.

\bibitem{Creminelli:2013nua}
P.~Creminelli, J.~Gleyzes, L.~Hui, M.~Simonovi\'c, and F.~Vernizzi,
  ``{Single-Field Consistency Relations of Large Scale Structure. Part III:
  Test of the Equivalence Principle},''
  \href{http://dx.doi.org/10.1088/1475-7516/2014/06/009}{{\em JCAP} {\bf 1406}
  (2014)  009},
\href{http://arxiv.org/abs/1312.6074}{{\tt arXiv:1312.6074 [astro-ph.CO]}}.

\bibitem{Horn:2014rta}
B.~Horn, L.~Hui, and X.~Xiao, ``{Soft-Pion Theorems for Large Scale
  Structure},'' \href{http://dx.doi.org/10.1088/1475-7516/2014/09/044}{{\em
  JCAP} {\bf 1409} (2014) no.~09, 044},
\href{http://arxiv.org/abs/1406.0842}{{\tt arXiv:1406.0842 [hep-th]}}.

\bibitem{Horn:2015dra}
B.~Horn, L.~Hui, and X.~Xiao, ``{Lagrangian space consistency relation for
  large scale structure},''
  \href{http://dx.doi.org/10.1088/1475-7516/2015/09/068}{{\em JCAP} {\bf 1509}
  (2015) no.~09, 068},
\href{http://arxiv.org/abs/1502.06980}{{\tt arXiv:1502.06980 [hep-th]}}.

\bibitem{Hui:2016ffo}
L.~Hui, ``{Symmetries in Large Scale Structure},''
{\em PoS} {\bf FFP14} (2016)  082.

\bibitem{Bordin:2016ruc}
L.~Bordin, P.~Creminelli, M.~Mirbabayi, and J.~Noreña, ``{Tensor Squeezed
  Limits and the Higuchi Bound},''
\href{http://arxiv.org/abs/1605.08424}{{\tt arXiv:1605.08424 [astro-ph.CO]}}.

\bibitem{Strominger:2013lka}
A.~Strominger, ``{Asymptotic Symmetries of Yang-Mills Theory},''
  \href{http://dx.doi.org/10.1007/JHEP07(2014)151}{{\em JHEP} {\bf 07} (2014)
  151},
\href{http://arxiv.org/abs/1308.0589}{{\tt arXiv:1308.0589 [hep-th]}}.

\bibitem{Cachazo:2014fwa}
F.~Cachazo and A.~Strominger, ``{Evidence for a New Soft Graviton Theorem},''
\href{http://arxiv.org/abs/1404.4091}{{\tt arXiv:1404.4091 [hep-th]}}.

\bibitem{Larkoski:2014bxa}
A.~J. Larkoski, D.~Neill, and I.~W. Stewart, ``{Soft Theorems from Effective
  Field Theory},'' \href{http://dx.doi.org/10.1007/JHEP06(2015)077}{{\em JHEP}
  {\bf 06} (2015)  077},
\href{http://arxiv.org/abs/1412.3108}{{\tt arXiv:1412.3108 [hep-th]}}.

\bibitem{Avery:2015rga}
S.~G. Avery and B.~U.~W. Schwab, ``{Noether's second theorem and Ward
  identities for gauge symmetries},''
  \href{http://dx.doi.org/10.1007/JHEP02(2016)031}{{\em JHEP} {\bf 02} (2016)
  031},
\href{http://arxiv.org/abs/1510.07038}{{\tt arXiv:1510.07038 [hep-th]}}.

\bibitem{Weinberg:1965nx}
S.~Weinberg, ``{Infrared photons and gravitons},''
\href{http://dx.doi.org/10.1103/PhysRev.140.B516}{{\em Phys. Rev.} {\bf 140}
  (1965)  B516--B524}.

\bibitem{Bondi:1962px}
H.~Bondi, M.~G.~J. van~der Burg, and A.~W.~K. Metzner, ``{Gravitational waves
  in general relativity. 7. Waves from axisymmetric isolated systems},''
\href{http://dx.doi.org/10.1098/rspa.1962.0161}{{\em Proc. Roy. Soc. Lond.}
  {\bf A269} (1962)  21--52}.

\bibitem{Sachs:1962wk}
R.~K. Sachs, ``{Gravitational waves in general relativity. 8. Waves in
  asymptotically flat space-times},''
\href{http://dx.doi.org/10.1098/rspa.1962.0206}{{\em Proc. Roy. Soc. Lond.}
  {\bf A270} (1962)  103--126}.

\bibitem{Strominger:2013jfa}
A.~Strominger, ``{On BMS Invariance of Gravitational Scattering},''
  \href{http://dx.doi.org/10.1007/JHEP07(2014)152}{{\em JHEP} {\bf 07} (2014)
  152},
\href{http://arxiv.org/abs/1312.2229}{{\tt arXiv:1312.2229 [hep-th]}}.

\bibitem{He:2014laa}
T.~He, V.~Lysov, P.~Mitra, and A.~Strominger, ``{BMS supertranslations and
  Weinberg's soft graviton theorem},''
  \href{http://dx.doi.org/10.1007/JHEP05(2015)151}{{\em JHEP} {\bf 05} (2015)
  151},
\href{http://arxiv.org/abs/1401.7026}{{\tt arXiv:1401.7026 [hep-th]}}.

\bibitem{Mirbabayi:2016xvc}
M.~Mirbabayi and M.~Simonovi\'c, ``{Weinberg Soft Theorems from Weinberg
  Adiabatic Modes},''
\href{http://arxiv.org/abs/1602.05196}{{\tt arXiv:1602.05196 [hep-th]}}.

\bibitem{Weinberg:2003sw}
S.~Weinberg, ``{Adiabatic modes in cosmology},''
  \href{http://dx.doi.org/10.1103/PhysRevD.67.123504}{{\em Phys. Rev.} {\bf
  D67} (2003)  123504},
\href{http://arxiv.org/abs/astro-ph/0302326}{{\tt arXiv:astro-ph/0302326
  [astro-ph]}}.

\bibitem{Kehagias:2016zry}
A.~Kehagias and A.~Riotto, ``{BMS in Cosmology},''
  \href{http://dx.doi.org/10.1088/1475-7516/2016/05/059}{{\em JCAP} {\bf 1605}
  (2016) no.~05, 059},
\href{http://arxiv.org/abs/1602.02653}{{\tt arXiv:1602.02653 [hep-th]}}.

\bibitem{Brown:1986nw}
J.~D. Brown and M.~Henneaux, ``{Central Charges in the Canonical Realization of
  Asymptotic Symmetries: An Example from Three-Dimensional Gravity},''
\href{http://dx.doi.org/10.1007/BF01211590}{{\em Commun. Math. Phys.} {\bf 104}
  (1986)  207--226}.

\bibitem{Strominger:1997eq}
A.~Strominger, ``{Black hole entropy from near horizon microstates},''
  \href{http://dx.doi.org/10.1088/1126-6708/1998/02/009}{{\em JHEP} {\bf 02}
  (1998)  009},
\href{http://arxiv.org/abs/hep-th/9712251}{{\tt arXiv:hep-th/9712251
  [hep-th]}}.

\bibitem{Strominger:2001pn}
A.~Strominger, ``{The dS / CFT correspondence},''
  \href{http://dx.doi.org/10.1088/1126-6708/2001/10/034}{{\em JHEP} {\bf 10}
  (2001)  034},
\href{http://arxiv.org/abs/hep-th/0106113}{{\tt arXiv:hep-th/0106113
  [hep-th]}}.

\bibitem{Brown:1992br}
J.~D. Brown and J.~W. York, Jr., ``{Quasilocal energy and conserved charges
  derived from the gravitational action},''
  \href{http://dx.doi.org/10.1103/PhysRevD.47.1407}{{\em Phys. Rev.} {\bf D47}
  (1993)  1407--1419},
\href{http://arxiv.org/abs/gr-qc/9209012}{{\tt arXiv:gr-qc/9209012 [gr-qc]}}.

\bibitem{Spradlin:2001pw}
M.~Spradlin, A.~Strominger, and A.~Volovich, ``{Les Houches lectures on de
  Sitter space},'' in {\em {Unity from duality: Gravity, gauge theory and
  strings. Proceedings, NATO Advanced Study Institute, Euro Summer School, 76th
  session, Les Houches, France, July 30-August 31, 2001}}, pp.~423--453.
\newblock 2001.
\newblock
\href{http://arxiv.org/abs/hep-th/0110007}{{\tt arXiv:hep-th/0110007
  [hep-th]}}.
\newblock

\bibitem{Balasubramanian:1999re}
V.~Balasubramanian and P.~Kraus, ``{A Stress tensor for Anti-de Sitter
  gravity},'' \href{http://dx.doi.org/10.1007/s002200050764}{{\em Commun. Math.
  Phys.} {\bf 208} (1999)  413--428},
\href{http://arxiv.org/abs/hep-th/9902121}{{\tt arXiv:hep-th/9902121
  [hep-th]}}.

\bibitem{deHaro:2000vlm}
S.~de~Haro, S.~N. Solodukhin, and K.~Skenderis, ``{Holographic reconstruction
  of space-time and renormalization in the AdS / CFT correspondence},''
  \href{http://dx.doi.org/10.1007/s002200100381}{{\em Commun. Math. Phys.} {\bf
  217} (2001)  595--622},
\href{http://arxiv.org/abs/hep-th/0002230}{{\tt arXiv:hep-th/0002230
  [hep-th]}}.

\bibitem{Balasubramanian:2001nb}
V.~Balasubramanian, J.~de~Boer, and D.~Minic, ``{Mass, entropy and holography
  in asymptotically de Sitter spaces},''
  \href{http://dx.doi.org/10.1103/PhysRevD.65.123508}{{\em Phys. Rev.} {\bf
  D65} (2002)  123508},
\href{http://arxiv.org/abs/hep-th/0110108}{{\tt arXiv:hep-th/0110108
  [hep-th]}}.

\bibitem{Ginsparg:1988ui}
P.~H. Ginsparg, ``{Applied Conformal Field Theory},'' in {\em {Les Houches
  Summer School in Theoretical Physics: Fields, Strings, Critical Phenomena Les
  Houches, France, June 28-August 5, 1988}}.
\newblock 1988.
\newblock
\href{http://arxiv.org/abs/hep-th/9108028}{{\tt arXiv:hep-th/9108028
  [hep-th]}}.
\newblock

\bibitem{Park:1998qk}
M.-I. Park, ``{Statistical entropy of three-dimensional Kerr-de Sitter
  space},'' \href{http://dx.doi.org/10.1016/S0370-2693(98)01119-8}{{\em Phys.
  Lett.} {\bf B440} (1998)  275--282},
\href{http://arxiv.org/abs/hep-th/9806119}{{\tt arXiv:hep-th/9806119
  [hep-th]}}.

\bibitem{Banados:1998gg}
M.~Banados, ``{Three-dimensional quantum geometry and black holes},''
  \href{http://arxiv.org/abs/hep-th/9901148}{{\tt arXiv:hep-th/9901148
  [hep-th]}}.
[AIP Conf. Proc.484,147(1999)].

\bibitem{Hinterbichler:2015pqa}
K.~Hinterbichler and A.~Joyce, ``{Hidden symmetry of the Galileon},''
  \href{http://dx.doi.org/10.1103/PhysRevD.92.023503}{{\em Phys. Rev.} {\bf
  D92} (2015) no.~2, 023503},
\href{http://arxiv.org/abs/1501.07600}{{\tt arXiv:1501.07600 [hep-th]}}.

\bibitem{Nicolis:2008in}
A.~Nicolis, R.~Rattazzi, and E.~Trincherini, ``{The Galileon as a local
  modification of gravity},''
  \href{http://dx.doi.org/10.1103/PhysRevD.79.064036}{{\em Phys. Rev.} {\bf
  D79} (2009)  064036},
\href{http://arxiv.org/abs/0811.2197}{{\tt arXiv:0811.2197 [hep-th]}}.

\bibitem{Hinterbichler:2014cwa}
K.~Hinterbichler and A.~Joyce, ``{Goldstones with Extended Shift Symmetries},''
  \href{http://dx.doi.org/10.1142/S0218271814430019}{{\em Int. J. Mod. Phys.}
  {\bf D23} (2014) no.~13, 1443001},
\href{http://arxiv.org/abs/1404.4047}{{\tt arXiv:1404.4047 [hep-th]}}.

\bibitem{Griffin:2014bta}
T.~Griffin, K.~T. Grosvenor, P.~Horava, and Z.~Yan, ``{Scalar Field Theories
  with Polynomial Shift Symmetries},''
  \href{http://dx.doi.org/10.1007/s00220-015-2461-2}{{\em Commun. Math. Phys.}
  {\bf 340} (2015) no.~3, 985--1048},
\href{http://arxiv.org/abs/1412.1046}{{\tt arXiv:1412.1046 [hep-th]}}.

\bibitem{Compere:2013bya}
G.~Compère, W.~Song, and A.~Strominger, ``{New Boundary Conditions for AdS3},''
  \href{http://dx.doi.org/10.1007/JHEP05(2013)152}{{\em JHEP} {\bf 05} (2013)
  152},
\href{http://arxiv.org/abs/1303.2662}{{\tt arXiv:1303.2662 [hep-th]}}.

\bibitem{Avery:2013dja}
S.~G. Avery, R.~R. Poojary, and N.~V. Suryanarayana, ``{An sl(2,$\mathbb{R}$)
  current algebra from $AdS_3$ gravity},''
  \href{http://dx.doi.org/10.1007/JHEP01(2014)144}{{\em JHEP} {\bf 01} (2014)
  144},
\href{http://arxiv.org/abs/1304.4252}{{\tt arXiv:1304.4252 [hep-th]}}.

\bibitem{Apolo:2014tua}
L.~Apolo and M.~Porrati, ``{Free boundary conditions and the AdS$_3$/CFT$_2$
  correspondence},'' \href{http://dx.doi.org/10.1007/JHEP03(2014)116}{{\em
  JHEP} {\bf 03} (2014)  116},
\href{http://arxiv.org/abs/1401.1197}{{\tt arXiv:1401.1197 [hep-th]}}.

\bibitem{Achucarro:1987vz}
A.~Achucarro and P.~K. Townsend, ``{A Chern-Simons Action for Three-Dimensional
  anti-De Sitter Supergravity Theories},''
\href{http://dx.doi.org/10.1016/0370-2693(86)90140-1}{{\em Phys. Lett.} {\bf
  B180} (1986)  89}.

\bibitem{Witten:1988hc}
E.~Witten, ``{(2+1)-Dimensional Gravity as an Exactly Soluble System},''
\href{http://dx.doi.org/10.1016/0550-3213(88)90143-5}{{\em Nucl. Phys.} {\bf
  B311} (1988)  46}.

\bibitem{Fitzpatrick:2014vua}
A.~L. Fitzpatrick, J.~Kaplan, and M.~T. Walters, ``{Universality of
  Long-Distance AdS Physics from the CFT Bootstrap},''
  \href{http://dx.doi.org/10.1007/JHEP08(2014)145}{{\em JHEP} {\bf 08} (2014)
  145},
\href{http://arxiv.org/abs/1403.6829}{{\tt arXiv:1403.6829 [hep-th]}}.

\end{thebibliography}\endgroup

\end{document}